\documentclass[useAMS,usenatbib,usegraphicx]{mn2e}

\usepackage{amsmath}
\usepackage{amsfonts}
\usepackage{amssymb}
\usepackage{subfigure}
\usepackage{color}

\newcommand{\aap}{A\&A} 
\newcommand{\mnras}{MNRAS} 
\newcommand{\apj}{ApJ} 
 
\newcommand{\apjs}{ApJS}
\newcommand{\aj}{AJ}
\newcommand{\pasp}{PASP} 
\newcommand{\nat}{Nat} 

\newcommand{\Msun}{\mbox{M}$_{\odot}$}

\begin{document}

\title[Dust lane galaxies trace gas-rich mergers]{Galaxy Zoo: dust lane early-type galaxies are tracers of recent, gas-rich minor mergers\thanks{This publication has been made possible by the participation of more than 250\,000 volunteers in the Galaxy Zoo 2 project. Their contributions are individually acknowledged at http://zoo2.galaxyzoo.org/authors}}

\author[Shabala et al.]{Stanislav S. Shabala,$^{1,2}$\thanks{E-mail: stanislav.shabala@utas.edu.au} Yuan-Sen Ting,$^{2,3}$ Sugata Kaviraj,$^{2,4}$ Chris Lintott,$^2$ 
\newauthor R. Mark Crockett,$^2$ Joseph Silk,$^2$ Marc Sarzi,$^5$ Kevin Schawinski,$^6$
\newauthor Steven P. Bamford$^7$ and Edd Edmondson$^8$\\
\\
$^1$ School of Mathematics \& Physics, University of Tasmania, Private Bag 37, Hobart 7001, Australia\\
$^2$ Oxford Astrophysics, Denys Wilkinson Building, Keble Road, Oxford OX1 3RH\\
$^3$ Ecole Polytechnique, 91128 Palaiseau Cedex, France\\
$^4$ Blackett Laboratory, Imperial College London, London SW7 2AZ\\
$^5$ Centre for Astrophysics Research, University of Hertfordshire, College Lane, Hatfield AL10 9AB\\
$^6$ Yale Center for Astronomy and Astrophysics, Yale University, PO Box 208121, New Haven, CT 06520, USA\\
$^7$ Centre for Astronomy \& Particle Theory, University of Nottingham, University Park, Nottingham NG7 2RD\\
$^8$ Institute of Cosmology and Gravitation, University of Portsmouth, Burnaby Road, Portsmouth PO1 3FX\\}

\date{Accepted 2012 January 19. Received 2011 December 27; in original form 2011 July 26}

\maketitle

\begin{abstract}

We present the second of two papers concerning the origin and evolution of local early-type galaxies exhibiting dust features. We use optical and radio data to examine the nature of active galactic nucleus (AGN) activity in these objects, and compare these with a carefully constructed control sample.

We find that dust lane early-type galaxies are much more likely to host emission-line AGN than the control sample galaxies. Moreover, there is a strong correlation between radio and emission-line AGN activity in dust lane early types, but not the control sample. Dust lane early-type galaxies show the same distribution of AGN properties in rich and poor environments, suggesting a similar triggering mechanism. By contrast, this is not the case for early types with no dust features. These findings strongly suggest that dust lane early-type galaxies are starburst systems formed in gas-rich mergers. Further evidence in support of this scenario is provided by enhanced star formation and black hole accretion rates in these objects. Dust lane early types therefore represent an evolutionary stage between starbursting and quiescent galaxies. In these objects, the AGN has already been triggered but has not as yet completely destroyed the gas reservoir required for star formation.

\end{abstract}


\section{Introduction}
\label{sec:introduction}

Dust lane early-type galaxies are those galaxies classified in Galaxy Zoo~2\footnote{http://zoo2.galaxyzoo.org} as early-type galaxies with dust lane features. The classification procedure and construction of the dust lane sample are described in detail in a companion paper, \citet[hereafter Paper~I]{kav12}. In that paper, we showed that dust lane early-type galaxies often exhibit morphological disturbances, show signs of recent star formation and have stellar ages that are older than starburst early types, but younger than the overall early-type population. We therefore argued that these objects are starburst systems, and suggested that they might arise from gas-rich minor mergers. While intergalactic medium accretion is also a plausible mechanism, we ruled it out in this case due to the presence of large amounts of dust, and the disturbed nature of most galaxies.

In the present paper we test this paradigm by examining the active galactic nucleus (AGN) properties of dust lane early-type galaxies. Numerous theoretical and observational arguments connect star formation and AGN triggering in galaxies. The basic point is that similar conditions (namely the presence of cold gas) are required for both a nuclear starburst and AGN fuelling. Once the supermassive black holes that reside at galaxy centres become active, radiative and kinetic feedback from these objects can profoundly affect their environments through heating and expulsion of gas. AGN feedback has been invoked to explain the cooling flow problem \citep{bin95}, and accounts for both the shape of bright end of the optical luminosity function \citep{sha09} and the strong correlation between black hole and bulge mass \citep{mag98,hae04} in the local Universe.

A number of diagnostics are used to identify AGN. Perhaps the best known is the so-called BPT \citep*{bal81} analysis, which relies on emission-line measurements to separate star-forming galaxies from AGN. Synchrotron emission from AGN jets and jet-inflated cocoons manifests itself at radio frequencies. X-ray and infrared diagnostics are also used.

Previous studies have shown that Seyfert and low-ionization nuclear emission-line region (LINER) elliptical hosts are more likely to contain dust than inactive ellipticals \citep{mar03,sim07}. However, it is unclear whether this finding can be extended to other AGN classifications. Recently, \citet{bes05b} showed that radio and emission-line AGN activities are largely independent at low radio luminosities, in contrast to the well-known correlation at high luminosities \citep{raw91}. This led \citet*{har07} to suggest an environmental origin for this dichotomy. Bright radio sources preferentially live in the field; these are also the objects that often exhibit emission-line AGN properties. By contrast, low-luminosity radio AGN tend to live in denser environments, and show little emission-line activity. \citeauthor{har07} argued that fundamentally different AGN fuelling mechanisms are likely to be at work in these two scenarios, with radio-loud objects in dense environments mainly fuelled by the relatively slow bremsstrahlung cooling out of the hot X-ray gas halo, while in poor environments mergers are primarily responsible for the sudden increase in cold gas content. The properties of the black hole accretion disc are different in the two cases. At high accretion rates associated with mergers, the disc assumes the standard geometrically thin, optically thick shape \citep[e.g.][]{sha73}; this disc is radiatively efficient and can also produce a radio jet. Thus, both emission-line and radio AGN are detected. At low accretion rates, however, the accretion disc puffs up and becomes radiatively inefficient, yielding a powerful radio jet but no corresponding emission lines \citep{mei01}. Therefore, a detailed analysis of both optical and radio AGN properties of our sample can help reconstruct the origins of the gas used in both star formation and AGN fuelling, and indicate whether dust lane early-type galaxies are indeed produced in galaxy interactions. Moreover, sizes and luminosities of radio AGN can be used to constrain their ages \citep{sha08} and therefore place these objects on an evolutionary sequence.

The homogeneity and size of our sample of dust lane early-type galaxies allow us to study the AGN-triggering mechanisms via a combination of photometry, spectroscopy and radio data. We create well-defined control samples, facilitating the construction of an evolutionary sequence and uncovering the significance of the observed dust deatures.

The paper is structured as follows. In Section~\ref{sec:SampleConstruction} we present our sample of dust lane early-type galaxies as well as a relevant control sample. In particular, we mimic the salient properties of dust lane galaxies in constructing the control sample. Section~\ref{sec:AGNselection} outlines the AGN selection procedure using two alternative definitions: radio luminosity and emission lines. We present our results in Section~\ref{sec:results}. Environmental dependence of dust lane AGN properties indicates that the origin of these objects is merger driven. We show that these mergers are necessarily recent, and gas-rich. Finally, we conclude in Section~\ref{sec:conclusions}.

Throughout the paper, we adopt the Hubble constant of $H_0=72$\,km\,s$^{-1}$Mpc$^{-1}$.

\section{Sample construction}
\label{sec:SampleConstruction}

\subsection{Dust lane early-type galaxies}
\label{sec:Sample_dustlanes}

\begin{figure*}
\centering
\subfigure[]{\includegraphics[height=0.2\textwidth,angle=0]{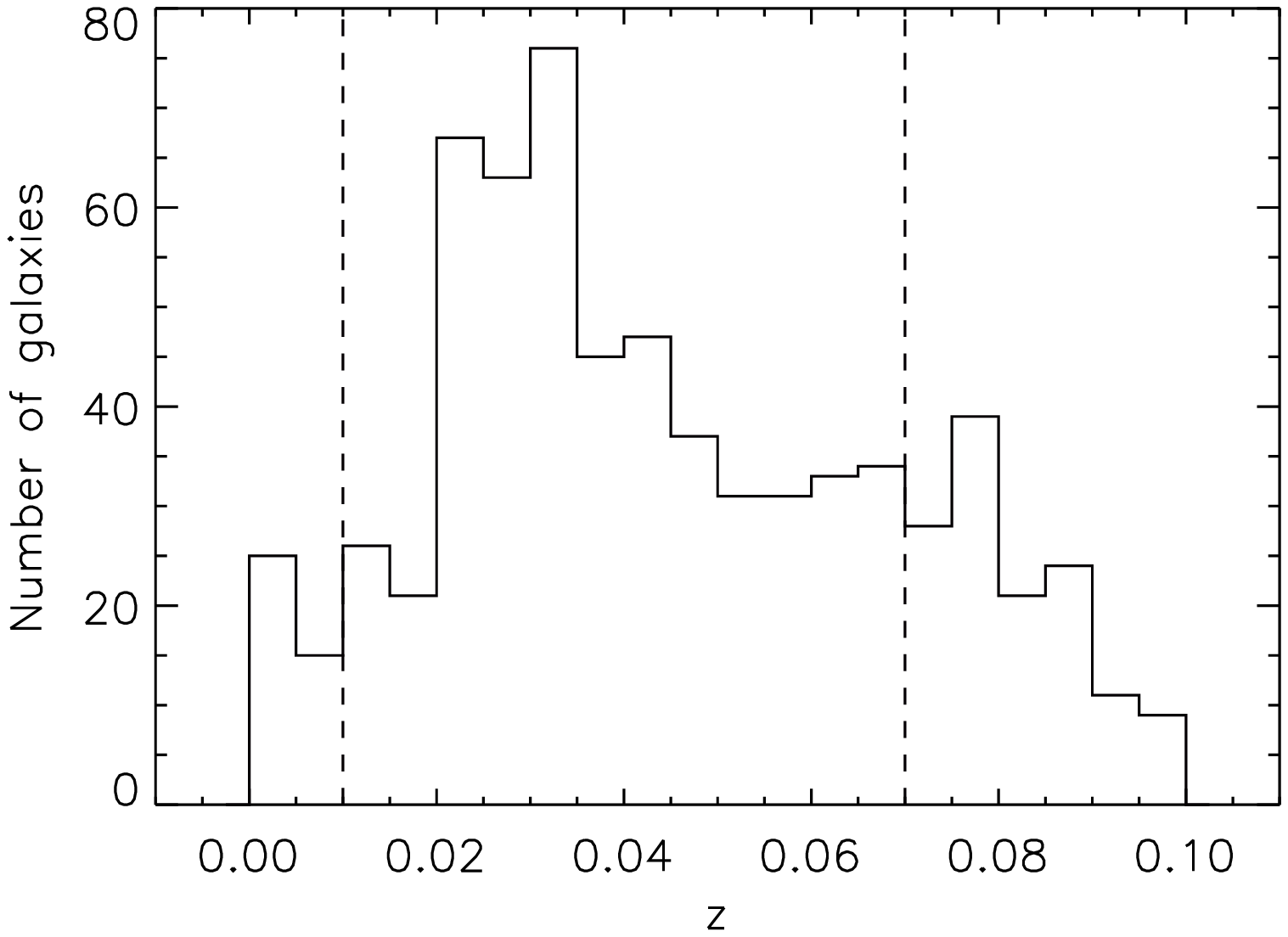}}
\subfigure[]{\includegraphics[height=0.2\textwidth,angle=0]{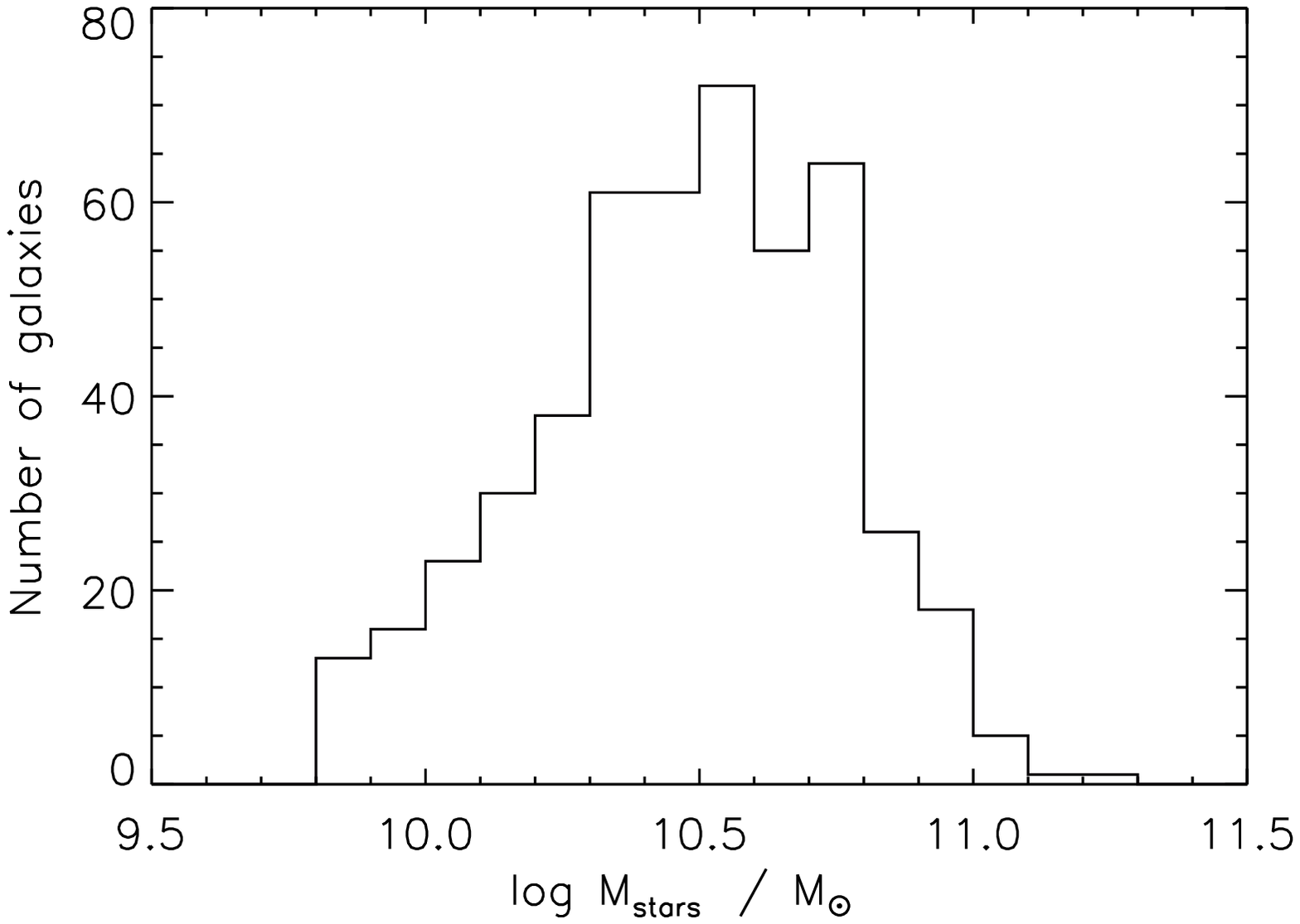}}
\subfigure[]{\includegraphics[height=0.2\textwidth,angle=0]{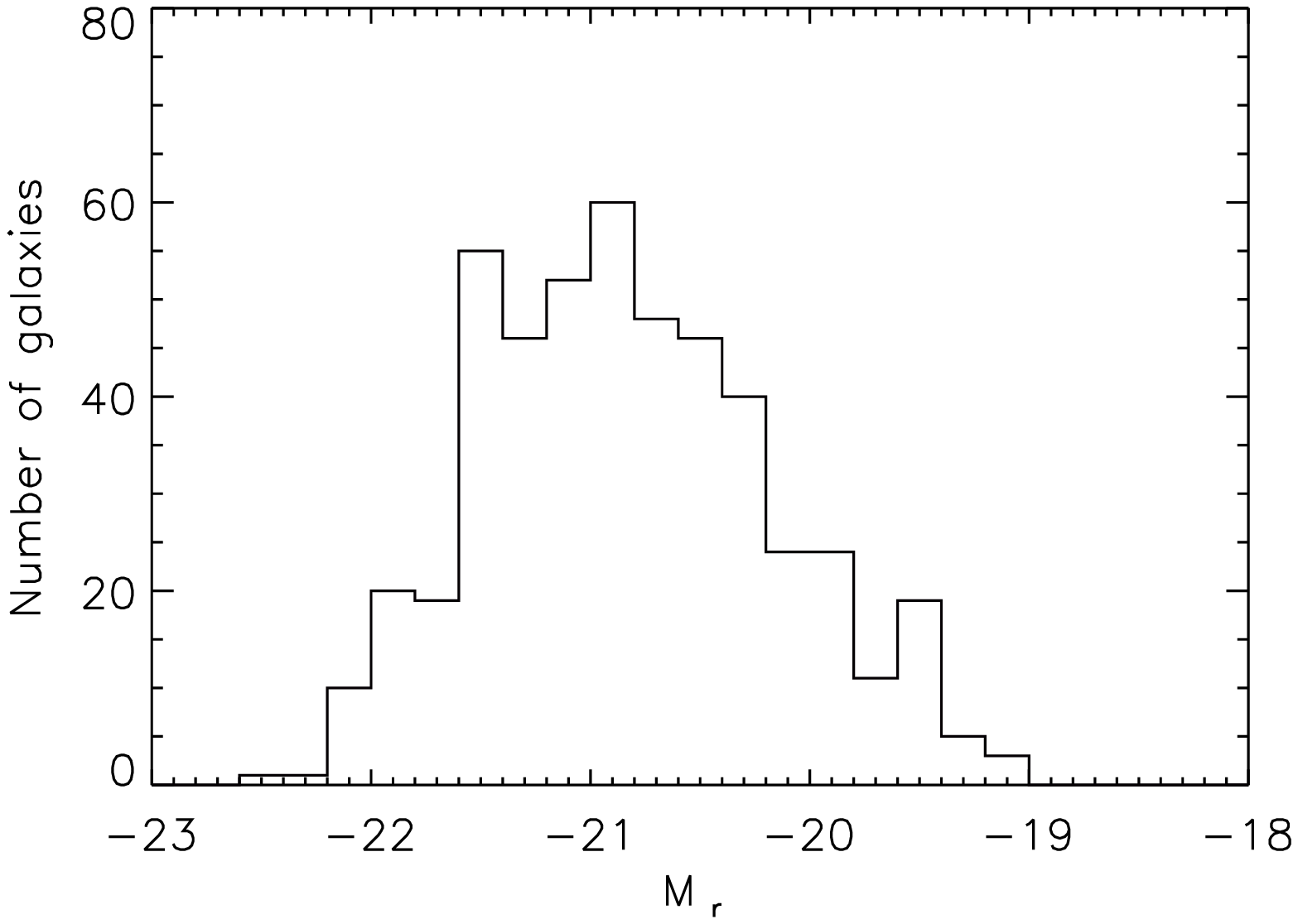}}
\caption{Redshift (left), stellar mass (middle) and $r$-band absolute magnitude (right) distribution of the dust lane early-type galaxy sample. The dashed lines show the adopted redshift cut.}
\label{fig:dustlanes_distributions}
\end{figure*}

The selection of early-type galaxies and a subsample of these exhibiting dust lane features from Sloan Digital Sky Survey (SDSS) DR7 is described in detail in Paper~I. The only slight difference between the samples analysed here and in Paper~I is the inclusion of bulge-dominated Sa galaxies in the present work, which were added to improve statistics. Sa galaxies constitute 23 per cent of the dust lane sample and exhibit very similar redshift distribution and disturbed fraction values to the rest of the sample. We made sure that their inclusion did not bias our conclusions by repeating the analysis without these objects, obtaining the same results. Fig.~\ref{fig:dustlanes_distributions} shows the stellar mass and redshift distribution of dust lane early-type galaxies.

\begin{table*}
\caption{Completeness and contamination as a function of SDSS-FIRST matching radius. The number of real matches is approximated by subtracting mock catalogue counts from the actual catalogue. A matching radius of 3 arcsec is adopted.}
\label{tab:catalogueMatching}
\centering
\begin{tabular}{lcccc}
\hline
\multicolumn{1}{l}{Separation} & Real matches & Completeness & False matches & Contamination \\ 
\multicolumn{1}{l}{(arcsec)}   &              & (per cent)   &               & (per cent)    \\
\hline
0.5                            & 1211         & 59.7         & 1             & 0             \\ 
1                              & 1742         & 85.9         & 1             & 0             \\ 
2                              & 1948         & 95.9         & 6             & 0.3           \\ 
3                              & 1974         & 96.8         & 12            & 0.6           \\ 
5                              & 2002         & 97.7         & 23            & 1.1           \\ 
15                             & 2174         & 98.9         & 170           & 8.4           \\ 
30                             & 2625         & 100          & 599           & 29.6          \\
\hline
\end{tabular}
\end{table*}

Our control sample consists of all early-type galaxies in the redshift range $0.01 \leq z \leq 0.07$. The upper limit on redshift corresponds to the maximum distance at which dust lanes can be reliably detected: as can be seen from Fig.~\ref{fig:dustlanes_distributions}, at redshifts $z \geq 0.08$, the fraction of detected dust lane features begins to drop, presumably due to the difficulty of identifying these with a naked eye. The low-redshift cut-off is imposed for two reasons. First, the radio-optical matching performed in the following section becomes increasingly inaccurate (as the same physical separation corresponds to a larger angular separation at lower redshift). Secondly, $V/V_{\rm max}$ corrections for Malmquist bias \citep[e.g.][]{sha08} to the radio luminosity function (RLF) become extreme if many low-luminosity (and hence low redshift) objects are included.

In principle, the variation in redshift and especially mass distribution between the dust lane and control samples can bias our conclusions. Therefore, for the control sample we mimic the redshift, stellar mass and absolute magnitude profiles of the relevant dust lane subsample.

\subsection{Radio-optical matching}
\label{sec:Sample_radioOptical}

We follow the procedure outlined in \citet{sha08} and \citet{bes05a} to construct radio-optical catalogues for dust lane early-type galaxies and the relevant control samples. Of the two large-area surveys at 1.4~GHz, Faint Images of the Radio Sky at Twenty Centimetres \citep*[FIRST;][]{bec95} provides much better positional accuracy than the NRAO VLA Sky Survey \citep[NVSS;][]{con98}. We paired sources from the FIRST catalogue with SDSS galaxies using a number of separation distances ranging from 1 to 30 arcsec. The point at which the number of matches stays approximately unchanged for any further increase in separation distance provides a good estimate of the best pairing distance estimate. This method yielded a best separation distance of a few arcseconds.

A different estimate is provided by generating a mock catalogue, constructed by offsetting all sources in the SDSS catalogue by 5 arcmin in right ascension. Both the mock and the actual SDSS catalogues were then paired with FIRST for a range of separation distances. At the largest separation distance, we expect all real matches to be captured, plus an unknown number of false matches (due to the allowed separation distance being too large). An approximate number of false matches was obtained by subtracting the number of total matches for the real catalogue from the corresponding value for the mock catalogue.

The best matching radius is obtained by maximizing the completeness, and minimizing contamination of the sample. Here, completeness is defined as the fraction of genuine optical-radio pairs being identified; and contamination as the number of false positives. Table~\ref{tab:catalogueMatching} led us to adopt a separation distance of 3 arcsec.

\citet{bes05a} found that many FIRST sources with offsets between 3 and 10 arcsec were found to be extended sources oriented close to the direction of the offset between the optical and radio positions. We followed these authors in refining the selection procedure to include FIRST sources which are offset from the SDSS position by more than 3 arcsec but less than 10 arcsec, and are oriented within 30$^\circ$ of the vector defined by the SDSS and FIRST coordinates. This resulted in an inclusion of an extra 20 sources ($1$ per cent of the sample) in the real catalogue, but only two ($0.1$ per cent) in the mock one, improving completeness without contaminating the sample.

We note that in the following analysis, NVSS maps (which are more sensitive to large-scale structure, but have less accurate position identifications) are used in parallel with FIRST for the galaxies identified as having a radio counterpart as outlined above. This was done to prevent any bias in observed radio AGN sizes and ages that would result from only using one of these surveys.

\section{AGN selection}
\label{sec:AGNselection}

\subsection{Infrared-radio correlation}
\label{sec:irRadio}

Most galaxies hosting radio AGN will also contain a contribution to radio luminosity from synchrotron emission associated with supernova-driven shocks. This needs to be subtracted from the total luminosity if the AGN component is to be defined properly. Using the relation of \citet*{yun01}, at 1.4~GHz the radio luminosity due to star formation is

\begin{equation}
 \frac{L_{\rm 1.4, SF}}{\rm W~Hz^{-1}} = 1.04 \times 10^{22} \left( \frac{\Psi}{\rm M_\odot yr^{-1}} \right).
\label{eqn:LradioSF}
\end{equation}

\begin{figure*}
\centering
\subfigure[]{\includegraphics[height=0.3\textwidth,angle=0]{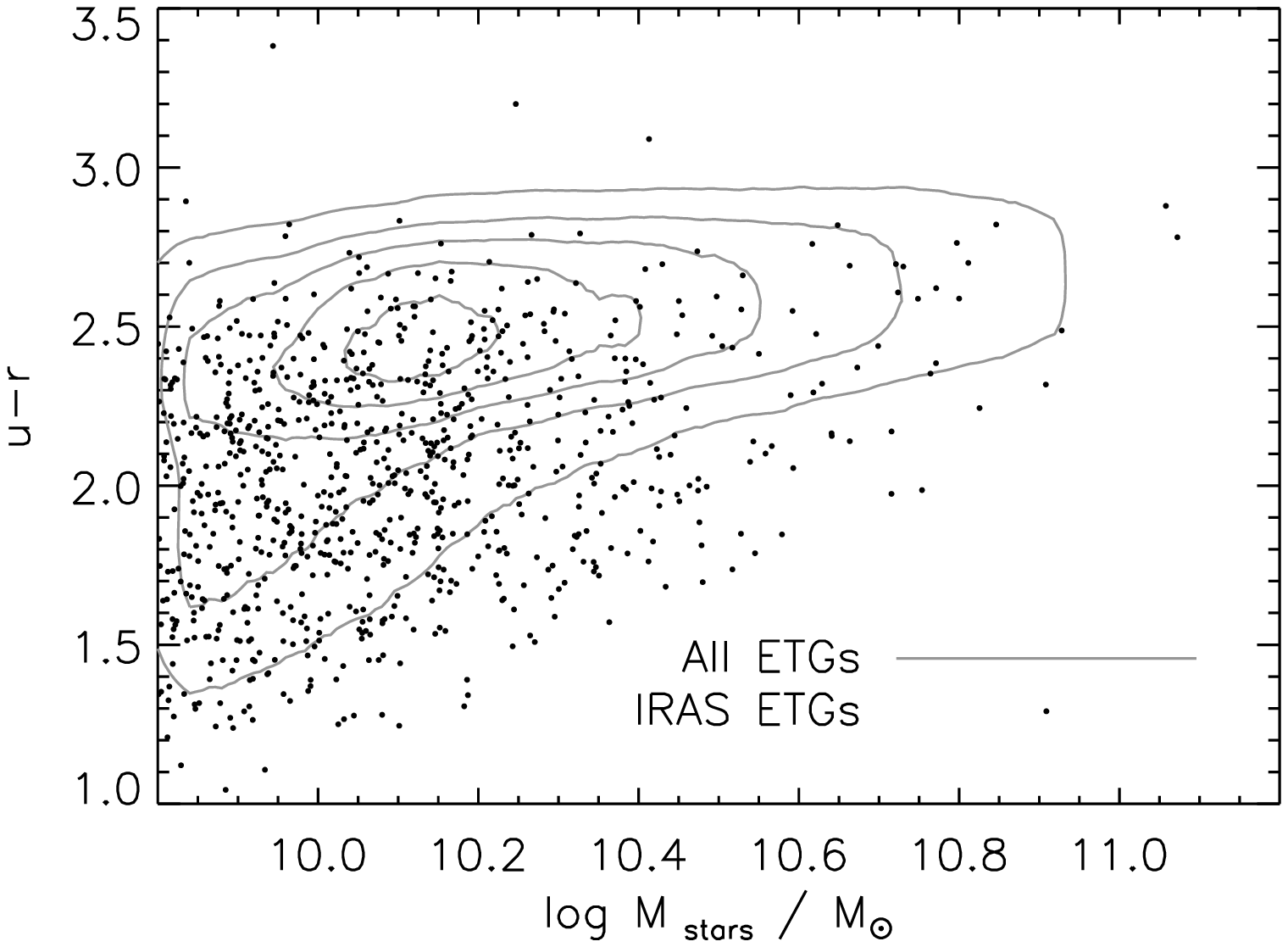}}
\subfigure[]{\includegraphics[height=0.3\textwidth,angle=0]{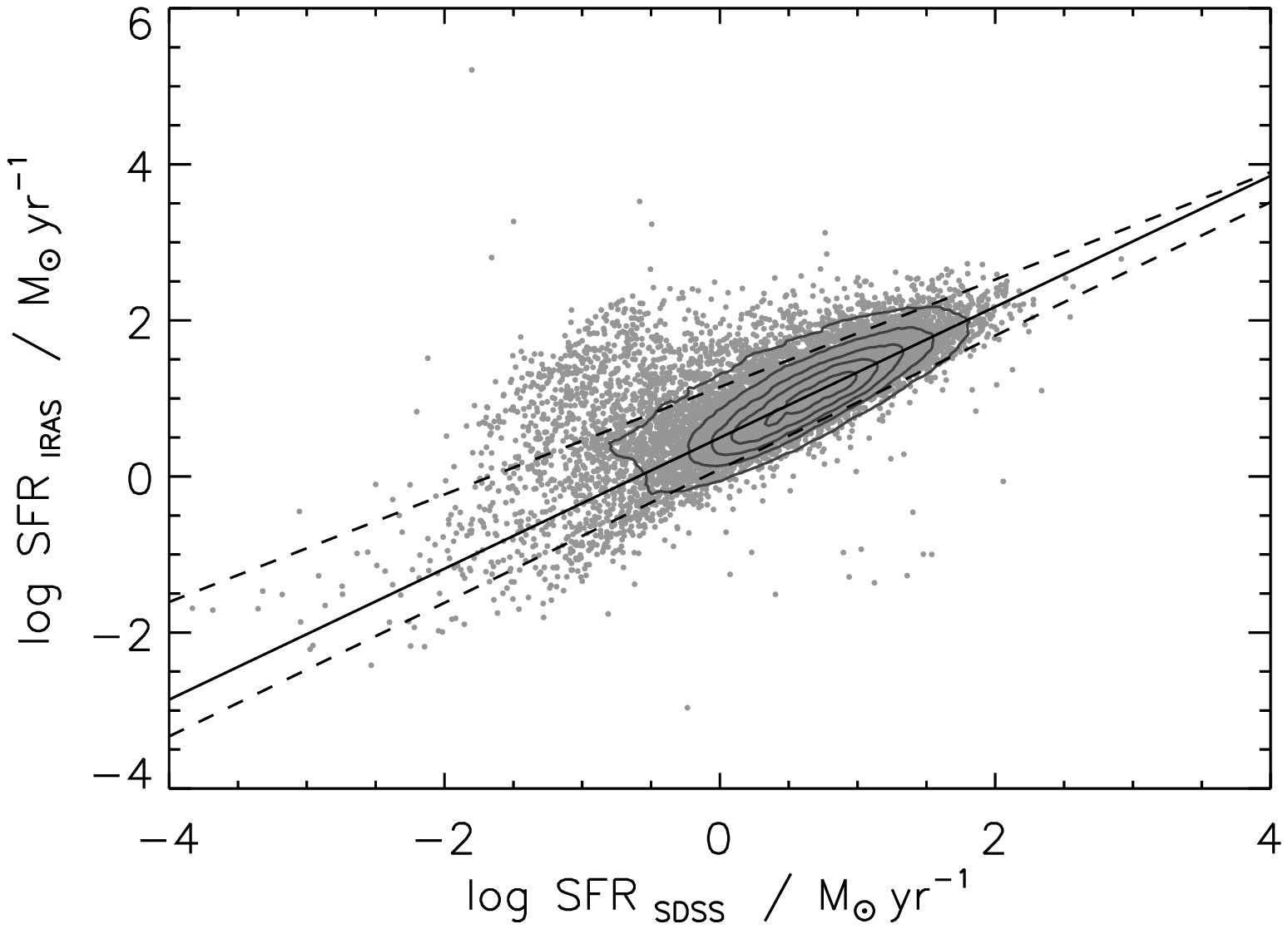}}
\caption{Left: colours of early-type galaxies with an {\it IRAS} counterpart (points), compared with all early-type galaxies (contours). {\it IRAS} selection is biased towards lower masses and bluer colours. Right: comparison of star formation rates derived from {\it IRAS} and SDSS. Contours and points represent the same data set of all SDSS galaxies with {\it IRAS} identifications. A good correlation is seen, with SDSS typically slightly underestimating the star formation rates. Solid line gives the median value, and dashed lines the 1.5$\sigma$ upper and lower bounds. In both plots the 10, 30, 50, 70, 90 per cent contours are shown.}
\label{fig:IRASvsSDSS}
\end{figure*}

Perhaps the best estimate of the star formation rate comes from far-infrared data, where it is unobscured. However, as Fig.~\ref{fig:IRASvsSDSS}(a) illustrates, selection of early-type galaxies with {\it IRAS} counterparts is biased towards blue early-type galaxies. A comparison of {\it IRAS} \citep{ken98} and SDSS-derived \citep{bri04,tre04} star formation rates for all galaxies in the redshift range of interest in Fig.~\ref{fig:IRASvsSDSS}(b) shows a good correlation between the two. The lower star formation rates in SDSS arise due to a combination of extinction and smaller aperture sizes. We therefore use the best-fitting relation to correct for the extinction associated with dust-corrected SDSS-derived star formation rates. Excess radio luminosities, which we attribute to AGN activity, are calculated by subtracting a star-formation contribution from the total luminosity. 

There is inherent scatter in the SFR-$L_{\rm 1.4, SF}$ relation given by equation~(\ref{eqn:LradioSF}), which can result in misclassification of star-forming galaxies as radio AGN. We therefore adopted a conservative requirement that, for an object to be classified as a radio AGN, the total radio luminosity must {\it significantly} exceed the mean value expected due to star formation alone. A number of cuts in radio luminosity were examined before adopting a 1.5$\sigma$ cut (i.e. $L_{\rm 1.4}$ must exceed $L_{\rm 1.4, SF}$ by more than 1.5 times the scatter in the {\it IRAS}-SDSS SFR relation). Taking a 1$\sigma$ cut results in classification of many blue early-type galaxies [as given by their ($u-r$) colours] as radio-loud AGN. A 2$\sigma$ cut yields the same results as our adopted cut, but with poorer statistics.

It is worth pointing out that here we are implicitly using nuclear star formation rates. As dust lane galaxies may exhibit more extended star formation than the control sample galaxies, considering only nuclear fluxes could in principle make dust lane hosts more likely to be classified as Seyferts or LINERs in the emission-line diagnostic in Section~\ref{sec:opticalAGN}. However, the dominant uncertainty of any such classification is much more likely to lie in the AGN/star-forming separation in the BPT plot rather than aperture effects. On the other hand, using nuclear fluxes ensures that sources identified as LINERs are genuine AGN rather than distributed old stellar populations.

This procedure identifies 65 radio-loud AGN (out of 484; 13 per cent) in the dust lane sample and 1188 out of 32\,590 (3.6 per cent) in the control sample of early-type galaxies.

\subsection{Emission line AGN}
\label{sec:opticalAGN}

An alternative way to identify AGN hosts is through the use of the emission-line diagnostic \citep[the so-called BPT diagram;][]{bal81}. We refer the reader to Paper~I for details of this process. Essentially, the 484 dust lane early-type galaxies and 32\,590 control sample of early types were classified according to their [O\,{\sc iii}]/H$\beta$ and [N\,{\sc ii}]/H$\alpha$ line ratios. Star-forming galaxies occupy a locus in the [O\,{\sc iii}]/H$\beta$ - [N\,{\sc ii}]/H$\alpha$ plane. Objects showing excess [O\,{\sc iii}] or [N\,{\sc ii}] emission but located close to the star-forming locus are classified as transition objects, i.e. galaxies which host both significant levels of star formation and AGN activity. Galaxies exhibiting even stronger [O\,{\sc iii}] or [N\,{\sc ii}] lines are classified as Seyfert AGN or LINERs. This analysis was performed for both the dust lane and control samples.

\begin{figure*}
\centering
\subfigure[]{\includegraphics[height=0.3\textwidth,angle=0]{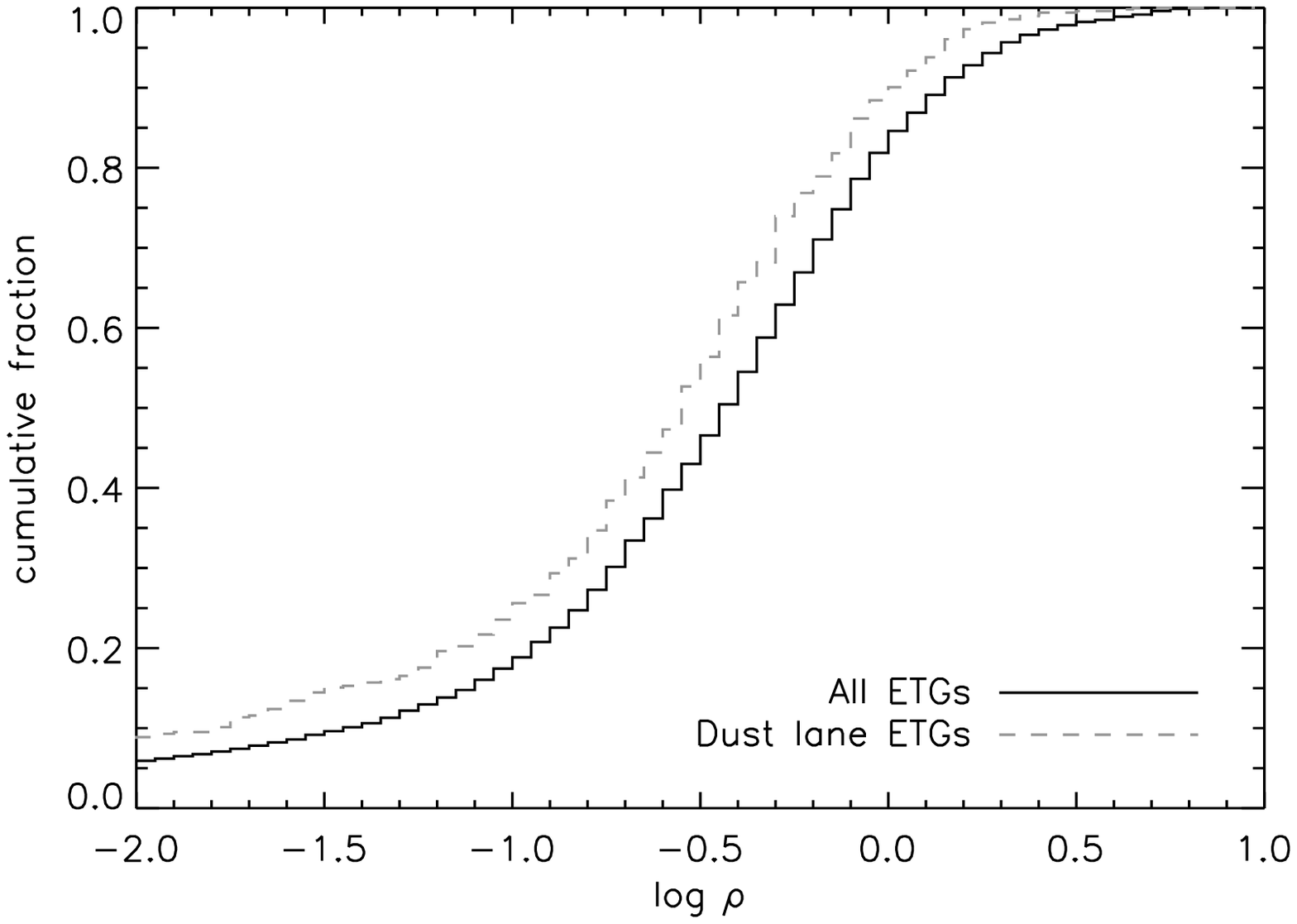}}
\subfigure[]{\includegraphics[height=0.3\textwidth,angle=0]{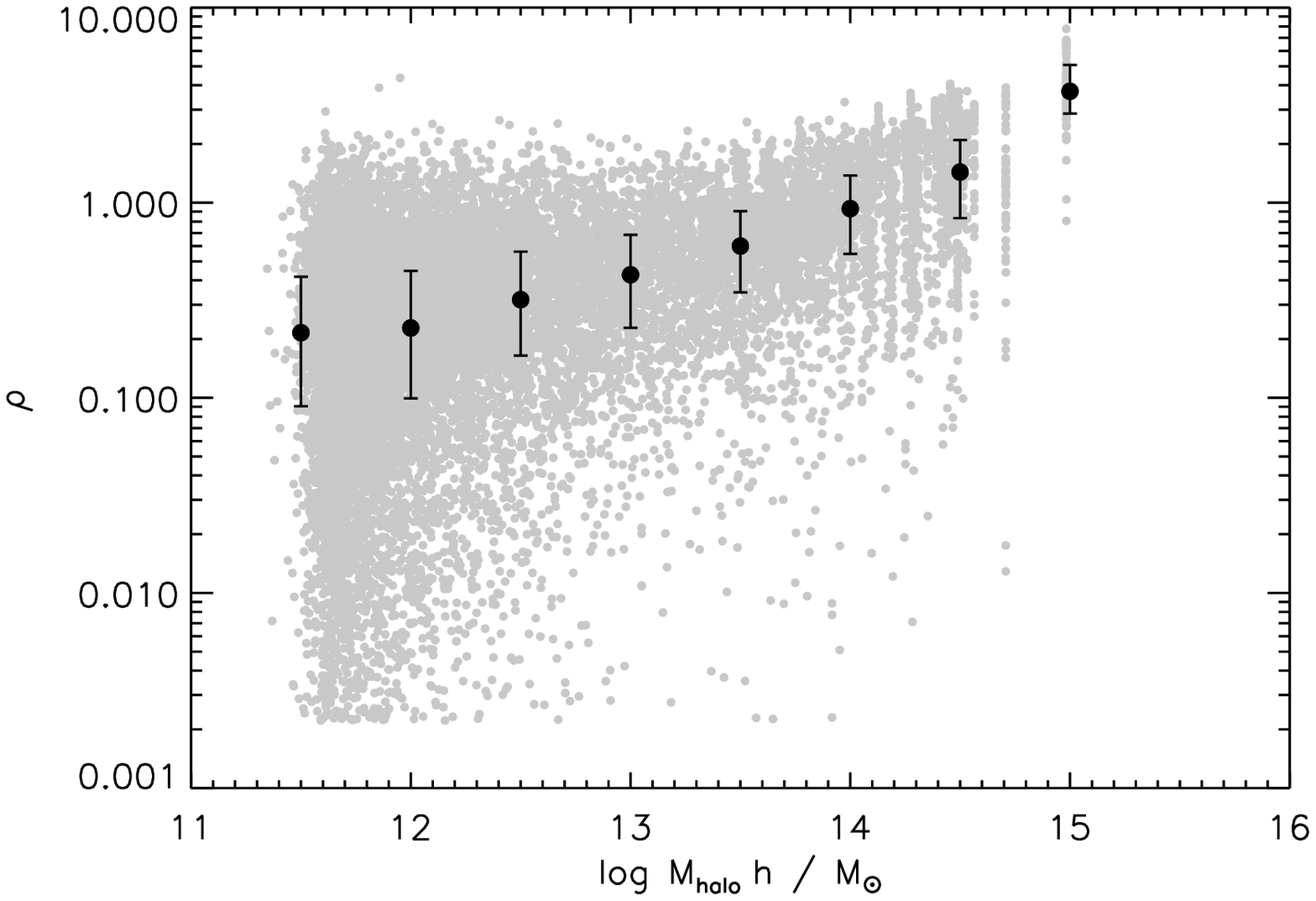}}
\caption{Left: distribution in environmental parameter $\rho$ as defined by \citet{sch07} for dust lane early-type galaxies (dashed) and a matched sample of all early-type galaxies (solid line) with $0.01<z<0.07$. Dust lane early-type galaxies reside in less dense environments than `average' early-type galaxies. Right: relationship between $\rho$ and the statistical density estimator of \citet{yan07}. Vertical `error bars' correspond to 25 and 75 percentile values.}
\label{fig:environments}
\end{figure*}

\section{AGN triggering and the AGN-starburst connection}
\label{sec:results}

\subsection{Environments}
As outlined in Section~\ref{sec:introduction}, a major motivation for considering two different modes of AGN activity is to probe both the AGN-triggering mechanisms and dust lane galaxy formation scenarios. Radio and emission-line AGN activities coincide for high-power radio sources, while low-luminosity radio AGN are identified with little optical activity. \citet{har07} suggested that such a scenario is consistent with high excitation line AGN being triggered by gas-rich mergers,\footnote{By `gas-rich' we do not necessarily mean a merger between two spiral galaxies. This is simply any merger which can supply a substantial amount of gas to the merger remnant over a short period of time. Cannibalism of dwarf galaxies and minor mergers with high merger ratios ($>1:10$) could plausibly meet this condition.} where large amounts of gas are added to the galaxy in a relatively short period of time, while radio-only AGN can be fuelled by cooling out of the hot halo gas \citep{bes05b,sha08}. In this scenario, a strong environmental dependence is expected, with merger-driven AGN being found in the field, and radio-only AGN residing in overdensities such as groups and clusters.

Fig.~\ref{fig:environments}(a) shows the distribution of environments for dust lane early-type galaxies and the matched control sample. In Paper~I, environment information was obtained by cross-matching our samples with the \citet{yan07} group catalogue. This statistical approach works well for large samples. In this paper, we quantify the nearest neighbour density within 2 Mpc using the environment parameter $\rho$ of \citet{sch07}. Fig.~\ref{fig:environments}(b) shows the strong correlation between the two quantities. In what follows, we adopt $\rho$ as a density indicator; however, all our findings are recovered when halo mass estimates of \citeauthor{yan07} are used instead. Dust lane early types are preferentially located in poor environments. The difference between the distributions in Fig.~\ref{fig:environments}(a), as given by a Kolmogorov-Smirnov test, is significant at the 99 per cent level.

It is worth noting that we are confident of having removed any potential mass dependence in the above environment result; nor is it subject to a selection bias. In all results described here, we made sure that for all control samples we mimicked the dust lane early-type mass, redshift and brightness distributions, as shown in Fig.~\ref{fig:dustlanes_distributions}. Given the strong evolution of many galaxy properties with mass \citep[e.g.][]{kau03,sha08}, this is crucial to ensuring that our findings are not simply a selection effect. Such careful construction of control samples is even more important in light of the visual selection employed in extracting dust lane early-type galaxies.

\subsection{Radio-loud fractions}

\begin{figure*}
\centering
\subfigure[Low-mass galaxies]{\includegraphics[height=0.3\textwidth,angle=0]{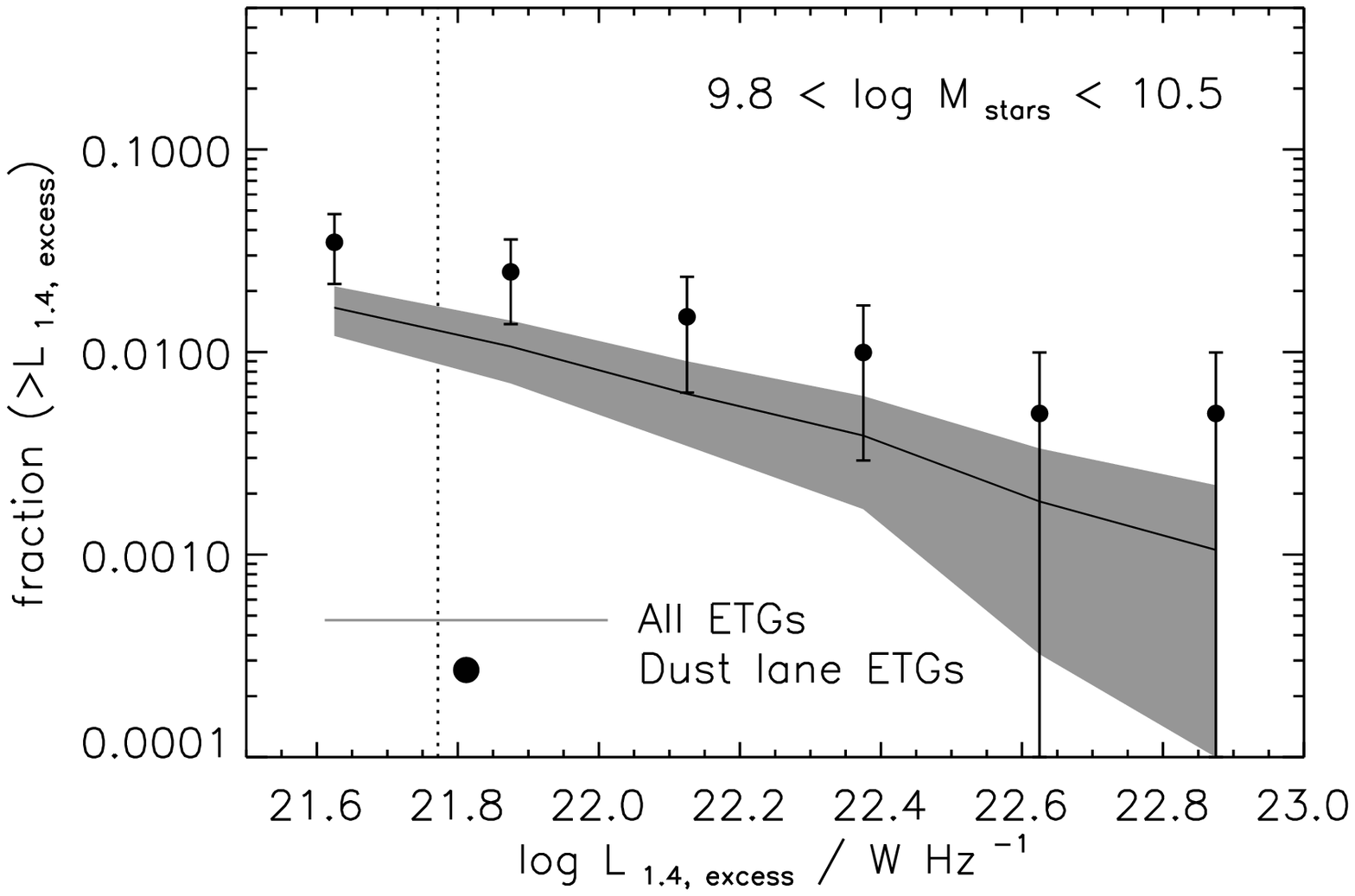}}
\subfigure[High-mass galaxies]{\includegraphics[height=0.3\textwidth,angle=0]{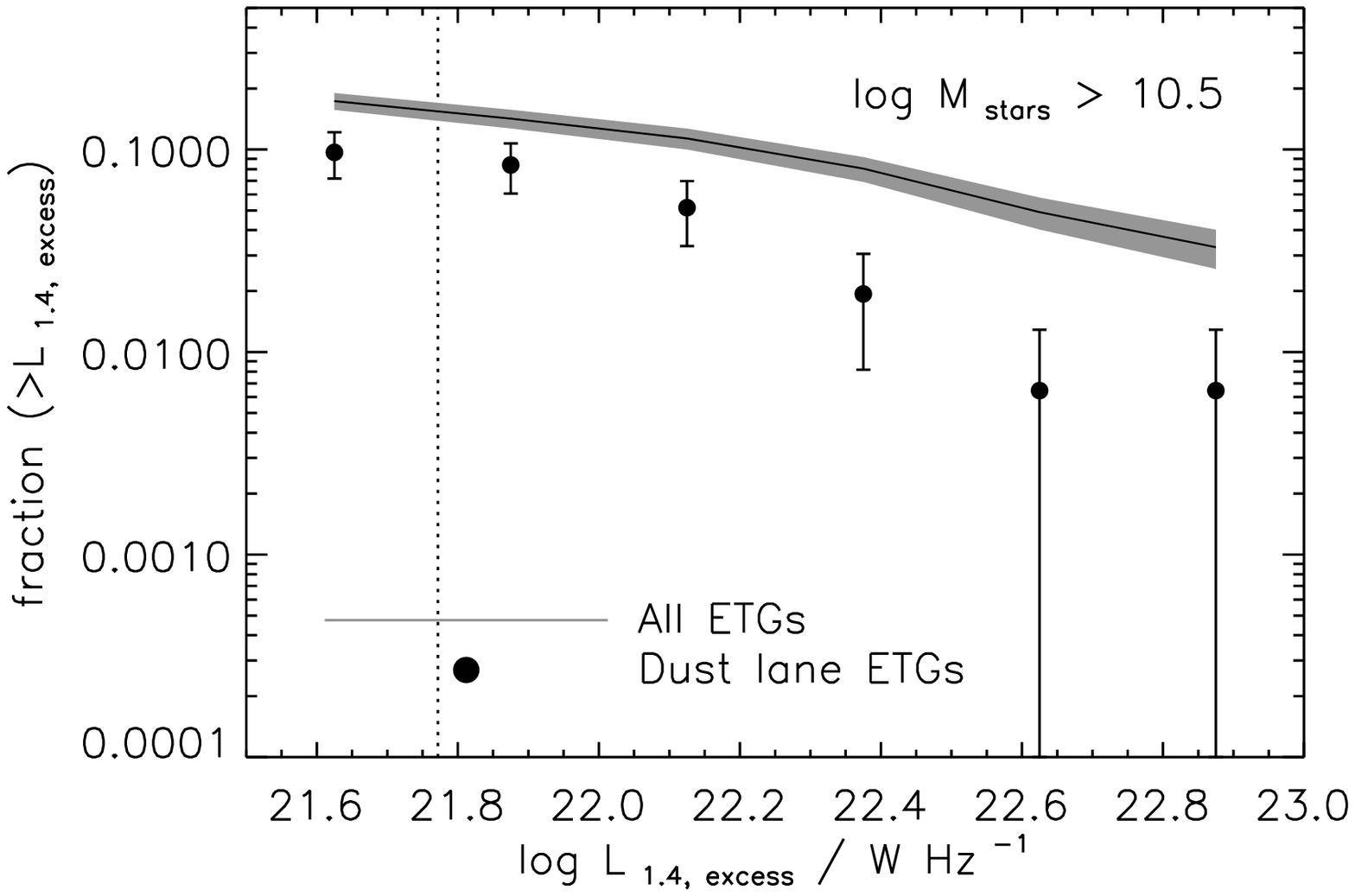}}
\caption{Left: cumulative fractional RLFs for low-mass dust lane (points) and a matched sample of low-mass early-type galaxies (shaded region) with $0.01<z<0.05$. Standard $V/V_{\rm max}$ corrections are applied. The dashed line corresponds to the completeness limit at the back of the volume. Right: same as the left-hand panel but for massive galaxies. Massive dust lane early types host significantly fewer radio AGN than the control sample.}
\label{fig:RLF_byMstars}
\end{figure*}

We now consider the radio AGN properties of early-type galaxies. Fig.~\ref{fig:RLF_byMstars} shows the cumulative fraction of galaxies hosting a radio AGN brighter than a given luminosity. The fractions were obtained by applying standard $V/V_{\rm max}$ corrections to radio source counts in order to correct for Malmquist bias. We adopted the redshift range $0.01 \leq z \leq 0.05$ for our samples in order for these corrections to not dominate genuine source counts. Since the colours and ages of local galaxies show a fairly well-defined bimodality, separating into the red sequence and blue cloud (with relatively few green valley objects) at a stellar mass of $\sim 3 \times 10^{10}$\Msun \citep{kau03}, we split our samples into high and low stellar mass bins at this value. Stellar masses, derived in Paper~I, are based on the calibrations of \citet{bel03}.

\begin{figure*}
\centering
\subfigure[Field]{\includegraphics[height=0.3\textwidth,angle=0]{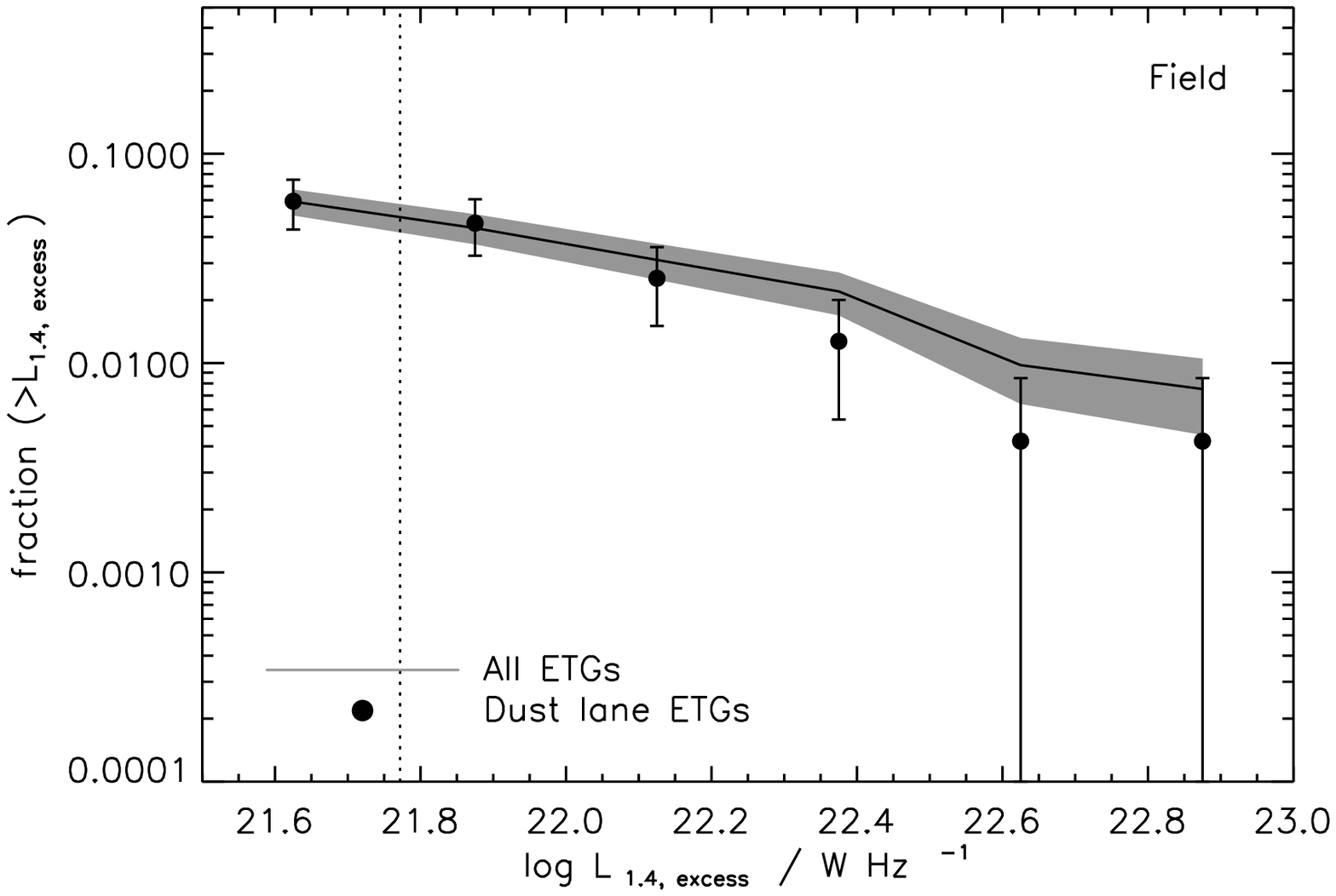}}
\subfigure[Group/Cluster]{\includegraphics[height=0.3\textwidth,angle=0]{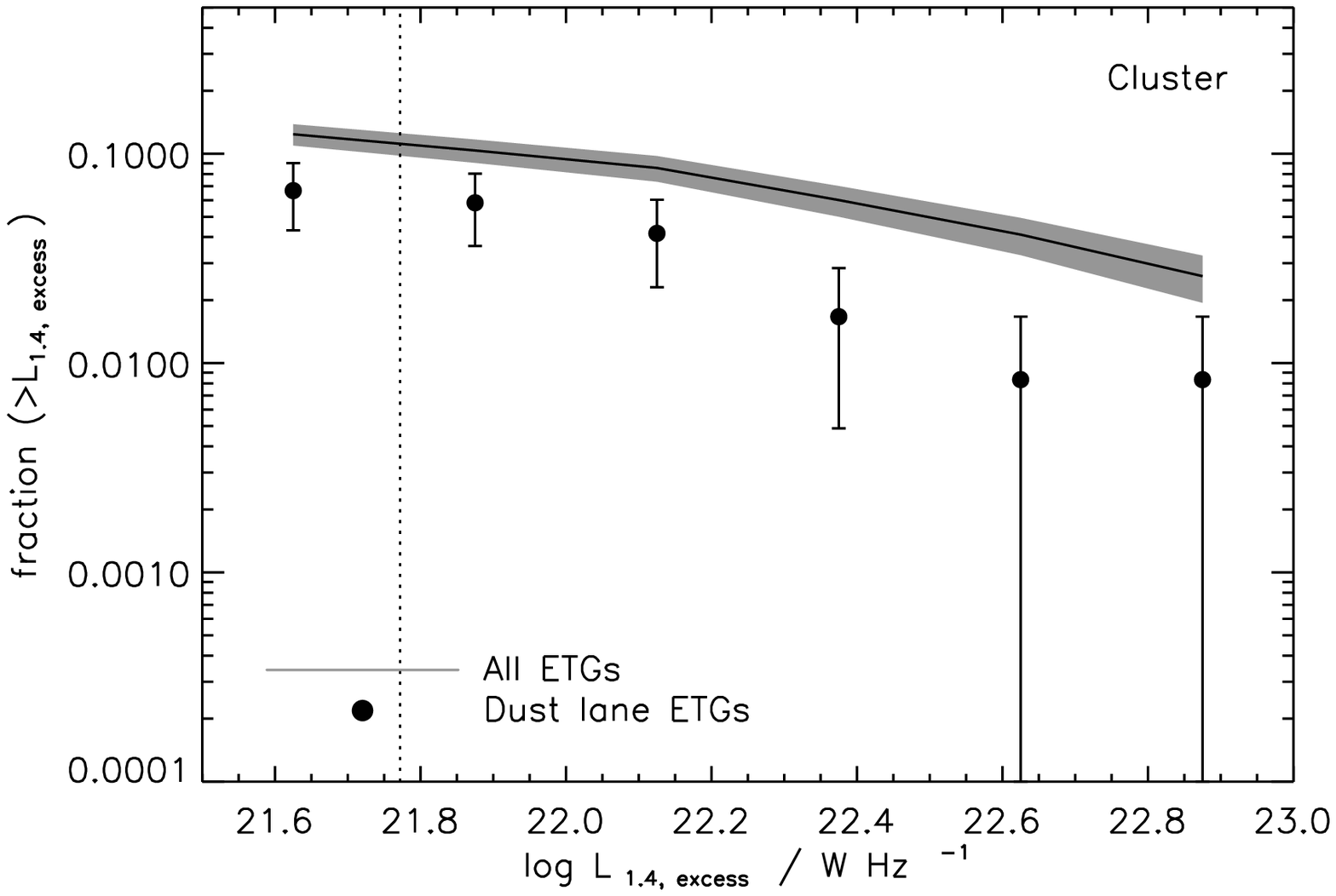}}
\caption{Same as Fig.~\ref{fig:RLF_byMstars} but for environments. Left: $\rho \leq 0.05$. Right: $\rho>0.05$. Dust lane early types have lower AGN fractions than the control sample in clusters, while there is no difference in the field.}
\label{fig:RLF_byEnvironment_dustlanesVscontrol}
\end{figure*}

At low masses, dust lane early-type galaxies show a marginally higher rate of AGN activity than the matched control sample of early-type galaxies, especially at lower luminosities. In massive galaxies, this result is reversed. One must be careful about the interpretation of these findings. In Fig.~\ref{fig:RLF_byEnvironment_dustlanesVscontrol} a similar result is obtained when the samples are split up by environment, rather than mass. Taking a cut at $\rho=0.05$, corresponding roughly to the transition between field and group/cluster environments \citep[][see also Fig.~\ref{fig:environments}b]{sch07}, dust lane and control sample early-type galaxies show comparable rates of AGN activity, while in groups/clusters the control sample of early types is more likely to host AGNs. Since massive early-type galaxies are preferentially found in clusters, the result of Fig.~\ref{fig:RLF_byMstars} is in fact an environment, rather than a mass, dependence. These findings are confirmed quantitatively by $\chi^2$ tests at 99 per cent level.

\begin{figure*}
\centering
\subfigure[Dust lane early types]{\includegraphics[height=0.3\textwidth,angle=0]{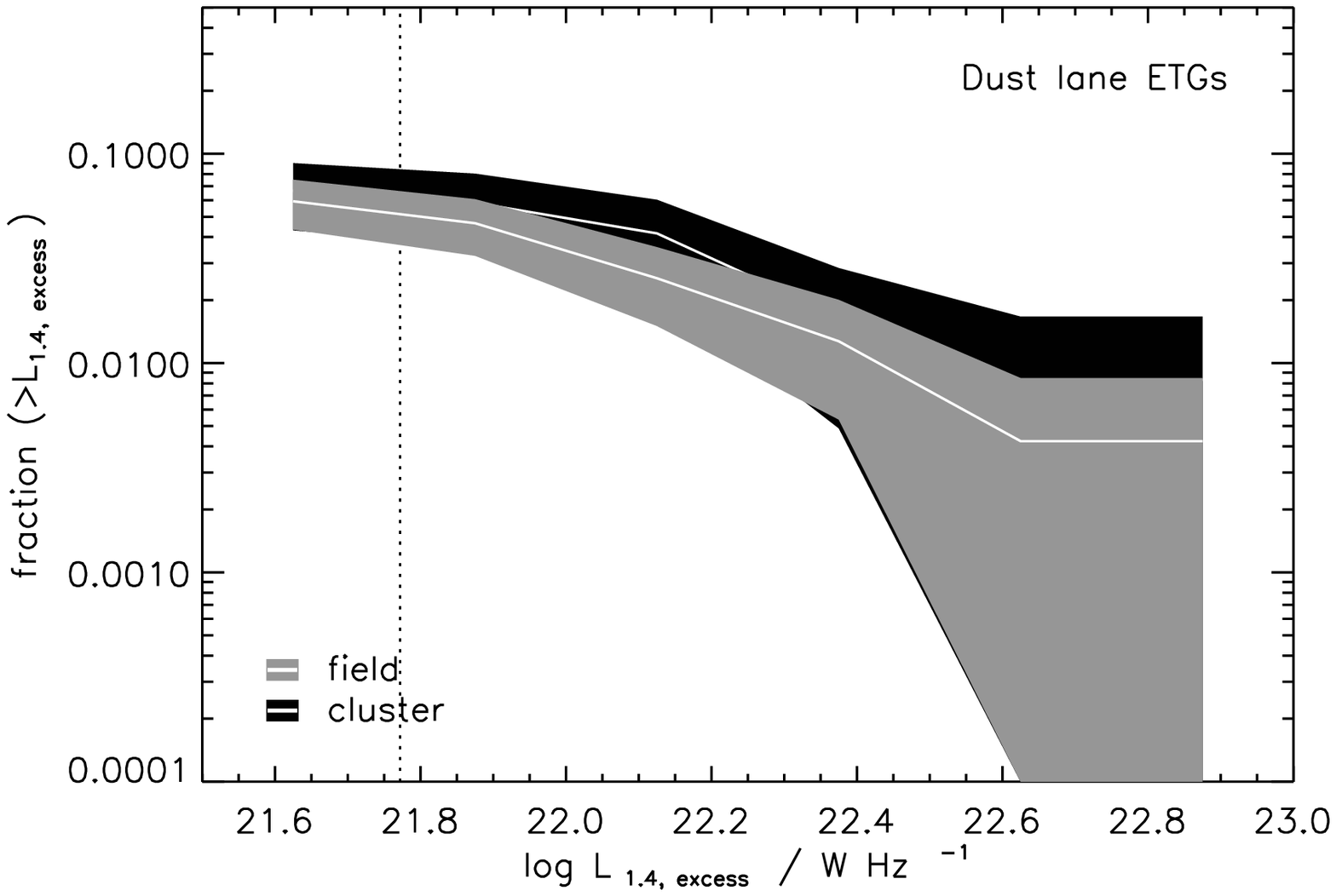}}
\subfigure[Control sample]{\includegraphics[height=0.3\textwidth,angle=0]{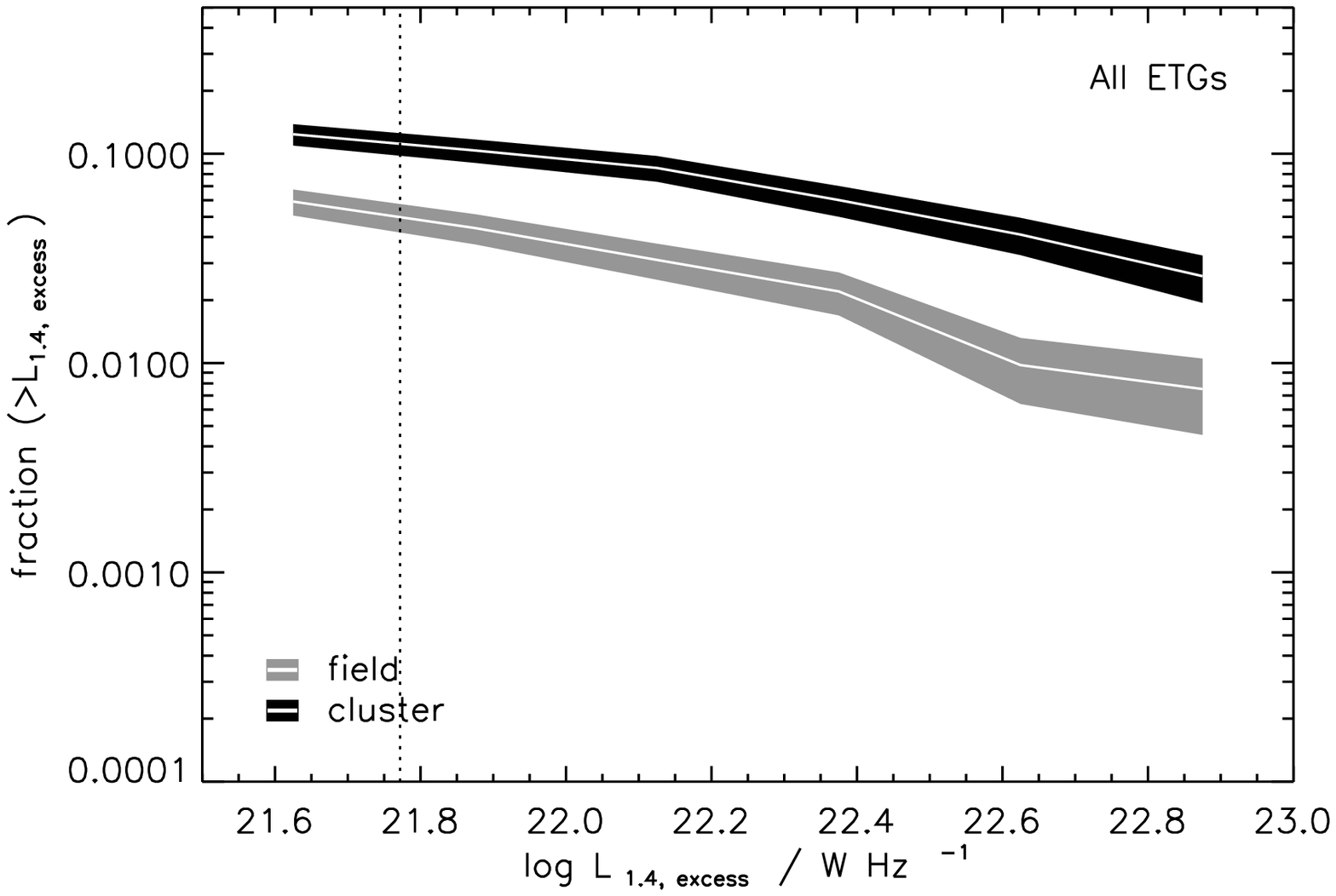}}
\caption{Same as Fig.~\ref{fig:RLF_byEnvironment_dustlanesVscontrol} but separated by environment for each of the two samples. Left: dust lane early-type galaxies. Right: control sample. In both cases, group/cluster galaxies are shown by the dark shaded region, and field galaxies as a light shaded region. There is no difference in radio AGN fractions between dust lane hosts in clusters and the field, while a significant enhancement in the cluster AGN fraction is seen in the control sample. This enhancement comes about due to a `cooling mode' of AGN fuelling available to massive early-type galaxies in clusters. In the field, the AGN fraction is set by the merger rate. The lack of AGN fraction dependence on environment for dust lane hosts strongly suggests that these objects trace gas-rich mergers.}
\label{fig:RLF_byEnvironment_separate}
\end{figure*}

\begin{table*}
\caption{Probability of finding an emission-line AGN given a radio-loud AGN. Entries denote the probability for the given galaxy type, while source counts are given in brackets. Dust lane early-type galaxies exhibit a strong correlation between emission-line and radio AGN activity.}
\label{tab:fracOpticalAGN_dustlanes}
\centering
\begin{tabular}{l|cc|c}
\hline
                      & log $(M/M_\odot)<10.5$ & log $(M/M_\odot) \geq 10.5$ & Total         \\
\hline
Dust lane early types & 0.83 (15/18)           & 0.92 (43/47)                & 0.89 (58/65)  \\
Control early types   & 0.53 (18/34)           & 0.25 (61/240)               & 0.29 (79/274) \\
\hline
\end{tabular}
\end{table*}

Interpretation of these results becomes easier when dust lane galaxies and the control sample of all early types are considered separately. As Fig.~\ref{fig:RLF_byEnvironment_separate} shows, the radio AGN fraction in dust lane early-type galaxies is independent of environment. Moreover, this fraction is also consistent with the radio AGN fraction of all early-type galaxies found in poor environments. In clusters this fraction rises significantly for early-type galaxies. Again, our conclusions are confirmed by $\chi^2$ tests at the 99 per cent level.

This latter result suggests the following interpretation. Radio AGN can be fuelled either by gas cooling out of the hot halo or a sudden supply of cold gas through a gas-rich merger event. The former scenario is more likely in massive early-type galaxies at the centres of clusters. Field/low-mass early types are less likely to be fuelled by this mechanism, with gas-rich mergers feeding the black hole in AGN hosts. Since in the local Universe such gas-rich merger events are much more rare than cooling episodes, cluster/high-mass early-type galaxies are more likely to host radio AGN than their field/low-mass counterparts.

The dust lane early-type population, both in the field and in clusters, shows the same distribution of radio properties as the field control population. Therefore, if mergers are the major trigger of radio AGN activity in all field early types, this is also the case for dust lane early-type galaxies, regardless of whether they are located in rich or poor environments. In other words, it appears as though a dust lane feature serves as a proxy for a recent gas-rich merger. This paradigm is supported by the finding in Paper~I that almost all dust lane galaxies exhibit morphological disturbances and recent starbursts.

According to this scenario, all dust lane early types should have undergone a recent gas-rich merger (and hence presumably host a detectable AGN), while only some of the early types in the control sample would have done the same. Figs~\ref{fig:RLF_byEnvironment_dustlanesVscontrol} and~\ref{fig:RLF_byEnvironment_separate}, however, show that the radio-loud fractions for these objects are similar. This is most likely due to our conservative (1.5$\sigma$ above emission due to star formation) radio luminosity cut adopted for AGN identification.

\begin{table}
\caption{Emission-line AGN fractions by type for dust lane early-type galaxies and a matched control sample of all early types with log~$(M_\star / M_\odot) <10.5$. Emission-line AGN in the second last row are defined as the galaxies classified as either transition objects, LINERs or Seyferts. Entries denote fractions of total population for the given galaxy type, while source counts are given in brackets. Note that the counts can be non-integral in the case of the matched control sample. Dust lane early types exhibit a much higher level of emission-line AGN activity than the control sample.}
\label{tab:fracAllAGN}
\begin{centering}
\begin{tabular}{l|cc} \hline
Classification    & Control     & Dust lane early types \\
\hline
Weak              & 0.78 (4510) & 0.12 (30)             \\
Star-forming      & 0.06 (368)  & 0.13 (32)             \\
Transition        & 0.06 (343)  & 0.28 (68)             \\
LINER             & 0.06 (369)  & 0.26 (62)             \\
Seyfert           & 0.03 (160)  & 0.21 (50)             \\
\hline
Emission-line AGN & 0.15 (874)  & 0.74 (180)            \\ 
Total             & 1.0 (5750)  & 1.0 (242)             \\
\hline
\end{tabular}
\end{centering}
\end{table}

\subsection{AGN triggering}
\label{sec:AGNtriggering}

One way the above paradigm can be tested is by comparing radio and emission-line AGN activity in dust lane early-type galaxies and the control sample. If gas-rich mergers are indeed responsible for fuelling the AGN in dust lane early types, one would expect a much higher fraction of radio AGN in these objects to also exhibit emission-line AGN activity, compared to objects in which gas cooling is the major AGN trigger. In Table~\ref{tab:fracOpticalAGN_dustlanes} we present the fraction of radio AGN hosts that are also identified as emission-line AGN. Almost all objects identified as radio AGN also show up as emission-line AGN in dust lane galaxies, but this is far from the case in the control sample, consistent with findings of the previous section.

Table~\ref{tab:fracAllAGN} shows the fraction of various AGN types, as defined in Paper~I, in dust lane early-type galaxies and the control sample. The control sample shows a much lower level of emission-line AGN activity ($\leq 20$ per cent) than dust lane early types ($>70$ per cent). Taken together with the above result that only dust lane galaxies (but not the control sample) show a significant correlation between emission-line and radio AGN activity, our interpretation is that dust lane early-type galaxies correspond to a phase in which radio and optical AGN activity is triggered.

\subsection{Stellar ages}

\begin{figure}
\centering
\includegraphics[height=0.3\textwidth,angle=0]{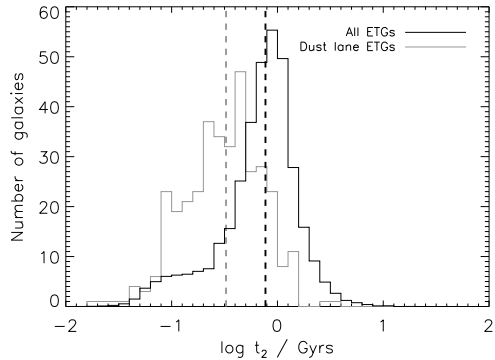}
\caption{Distribution of starburst ages for dust lane early-type galaxies and a matched control sample of all early types. Median ages are shown by dashed lines. Dust lane early-type galaxies host starbursts that are younger than those in all early-type galaxies.}
\label{fig:t2control}
\end{figure}

If dust lane early types are indeed associated with gas-rich mergers, an obvious question concerns their place in the evolution of a post-merger early-type galaxy. Following \citet{kav10} we estimate the age of the last starburst using the double colour $($NUV$-u)-(g-z)$, which is relatively dust insensitive. Fig.~\ref{fig:t2control} shows the starburst age distributions for the dust lane sample and a matched control sample of all early-type galaxies. The starburst ages of dust lane early-type galaxies are younger than the average early-type galaxy.

\subsection{Radio source ages}

\begin{figure*}
\centering
\subfigure[starburst ages]{\includegraphics[height=0.3\textwidth,angle=0]{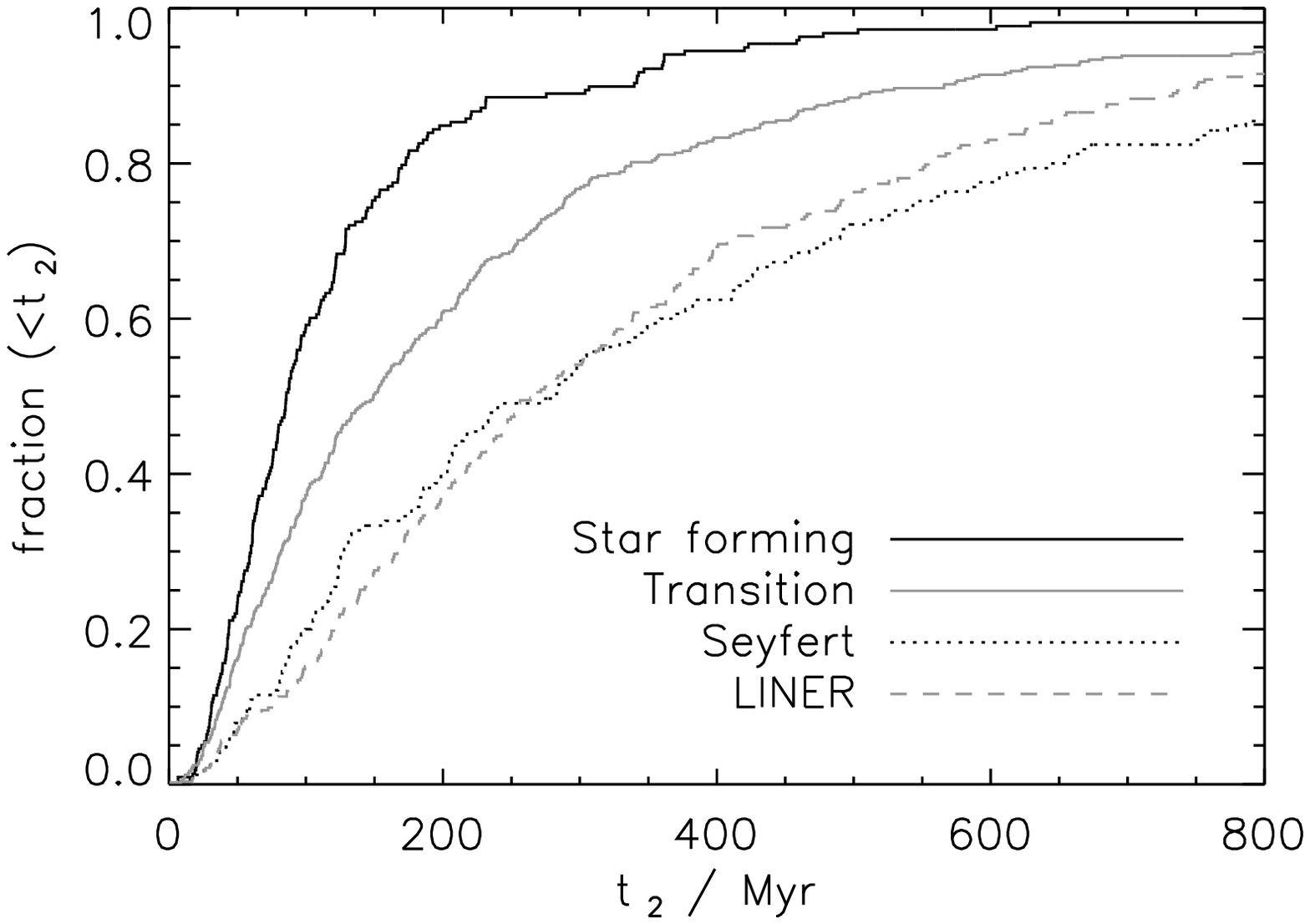}}
\subfigure[starburst and radio ages]{\includegraphics[height=0.3\textwidth,angle=0]{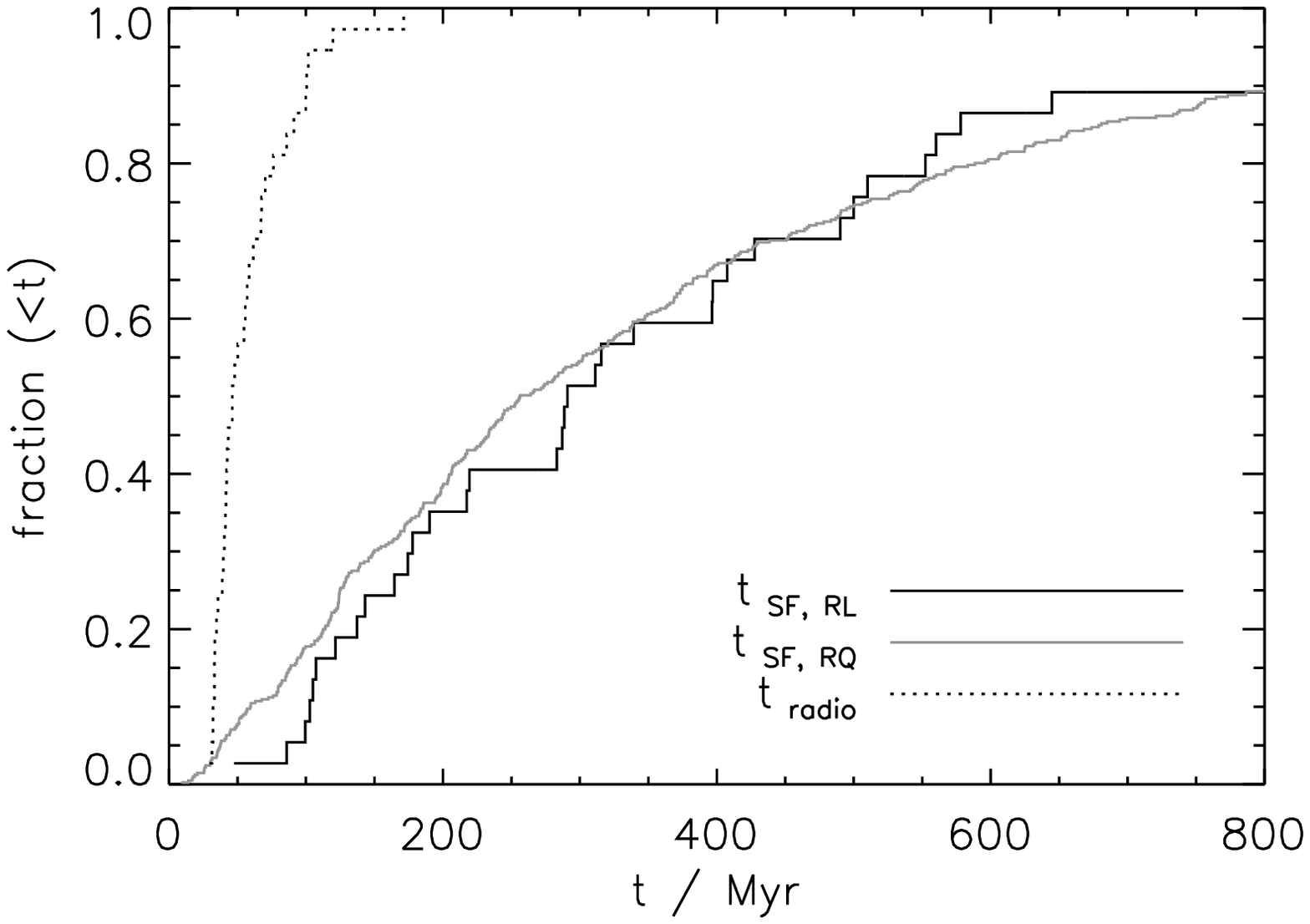}}
\caption{Left: distribution of starburst ages for various emission-line classifications of dust lane early-type galaxies. Both radio-loud and radio-quiet objects are included. An evolutionary sequence of starburst $\rightarrow$ transition object $\rightarrow$ Seyfert/LINER is seen. Right: comparison of starburst and radio source ages for dust lane early types. There is no difference in starburst age distributions between radio-loud and radio-quiet AGN.}
\label{fig:SFtimescales}
\end{figure*}

As shown in the previous section, dust lane early-type galaxies have stellar populations that are younger than the overall early-type population. Thus, these objects are likely to be starburst or post-starburst galaxies, observed during the transition between the blue and red-and-dead phases. The correlation between optical and radio AGN activity suggests gas-rich minor mergers (since major mergers are rare in the low-redshift Universe) may drive star formation and AGN activity. In this scenario, the central engine of the AGN is endowed with so much fuel that the accretion disc settles into a radiatively efficient, optically thick, geometrically thin state \citep[e.g.][]{mei01} and yields both an observable radio jet and narrow-line emission. Tables~\ref{tab:fracOpticalAGN_dustlanes} and \ref{tab:fracAllAGN} suggest this is not the case for the later (control early types) stages of evolution. This is likely to be because the merger-driven AGN activity has mostly expelled the available gas, and the galaxy migrated to a quiescent state. The mean starburst age for the control sample is $700$~Myr, which is much larger than both the transition time of $\sim 200$~Myr between the starburst and transition (i.e. starburst+AGN) phase argued by \citet[see also Fig.~\ref{fig:SFtimescales}]{sch07} and the maximum AGN lifetime of up to a few hundred Myr \citep{sha08}. Therefore, most of these objects are not expected to host AGN that would have a common origin with the starburst. Conversely, with their mean starburst ages of $\sim 300$~Myr, dust lane early types fall in the `sweet spot' of being old enough to host an AGN, but not so old that AGN activity has been terminated.

We test this paradigm by deriving radio AGN ages for those galaxies that were classified as having excess 1.4-GHz luminosity. This was done by generating a library of tracks in radio luminosity- linear size space. As discussed at length in \citet{sha08}, given a density profile for the atmosphere into which the radio source is expanding, and aspect ratio of the source, its luminosity and size allow jet power and age of the radio source to be determined uniquely. We use observations of local early-type galaxies \citep{all06} for the X-ray gas density profile, and adopt an axial ratio of $R_{\rm T}=2$. A number of caveats are associated with this analysis. Our assumption for the gas density profile may not be applicable for dust lane early-type galaxies, most of which show disturbed morphologies (Paper~I). Furthermore, hotspot contribution to radio luminosity can be comparable with that of the lobes at 1.4~GHz. However, \citet{sha08} estimate that the derived time-scales are accurate to better than a factor of two.

Importantly, the radio source models of \citet{kai97a} and \citet*{kai97b} are only applicable for edge-brightened [Fanaroff-Riley type II (FR II)] sources. On the other hand, most radio AGN in the local universe are core-dominated FR I objects. However, most of the resolved objects in our sample are FR II objects, for which the modelling is applicable. Only upper limits on ages can be placed for unresolved radio sources.

\begin{table*}
\caption{Fractional distribution of radio source sizes. There is no statistically significant difference between dust lane early-type galaxies and the control sample. Most radio sources in the dust lane sample are unresolved with FIRST, and only upper limits on their sizes are available.}
\label{tab:radioSizesAges}
\centering
\begin{tabular}{l|cc|c}
\hline
                      & size $\leq$ 30 kpc & size $>$ 30 kpc  & Total            \\
\hline
Dust lane early types & 0.16 (43/270)      & 0.008 (3/270)    & 0.17 (46/270)    \\
Control early types   & 0.058 (481/8951)   & 0.0007 (36/8951) & 0.058 (517/8951) \\
\hline
\end{tabular}
\end{table*}

Table~\ref{tab:radioSizesAges} gives the size distribution of radio AGN for the dust lanes and the matched control sample. The distributions are statistically indistinguishable at the 10 per cent level. However, the number of dust lane radio AGN in our sample (46) is small, and most of these are unresolved. High resolution imaging will be required to address this issue properly.

Fig.~\ref{fig:SFtimescales} compares the derived radio source ages with the ages of stellar populations for dust lane early-type galaxies. Seyferts and LINERs have similar age distributions, with median values around 300~Myr. There is no difference between radio-loud and radio-quiet emission-line AGN. We interpret this as evidence for the simultaneous triggering of the radio jet and line emission. Seyfert-like line emission will come from the radiatively efficient AGN disc, while LINER-like emission can either come from the disc or be distributed throughout the radio cocoon \citep[e.g.][]{kra11}. This picture is further supported by the $100-150$~Myr offset between radio and starburst ages in Fig.~\ref{fig:SFtimescales}(b) being consistent with the age difference between starburst and transition objects in Fig~\ref{fig:SFtimescales}(a).

Our interpretation of the origin and evolution of dust lane early-type galaxies is then as follows. A quiescent early-type galaxy undergoes a gas-rich minor merger. The resulting gas inflow triggers a starburst. Some time later (typically $<200$~Myr) the AGN switches on. Circumnuclear starbursts could be the reason for this delay, with powerful winds driven by OB stars and supernovae suppressing AGN fuelling temporarily \citep{dav07}. Alternatively, mass loss from newly formed stars could help fuel AGN activity, providing a possible mechanism of funneling gas towards the central engine. Once the AGN switches on, the accretion rate onto the central black hole is high enough (typically at least a few per cent of the Eddington value) for the accretion disc to be in a classical, radiatively efficient thin disc state, and the AGN is observed in both emission lines and at radio wavelengths. This phase is accompanied by the presence of significant amounts of dust, and the early-type galaxy is observed as having a dust lane feature. Eventually, AGN feedback heats up and/or expels the gas from the galaxy \citep[this is evidenced for example by the depletion of molecular gas reservoir on time-scales of around 200 Myr;][]{sch09}. Both the star formation and AGN activity are truncated, with the galaxy returning to a quiescent state.

\subsection{Origin of dust}

In Paper~I we showed that the mass of dust in the observed features is too great to come from stellar mass loss, suggesting strongly an external origin for this dust. Since dust and gas are typically coupled, one might expect dust lane early-type galaxies to be gas rich, consistent with the gas-rich merger origin of these objects. As shown in Paper~I, dust lane early types do show enhanced levels of star formation compared to a control sample of all early-type galaxies. However, Fig.~\ref{fig:t2control} shows that galaxies with dust features also have younger starburst ages, and it is therefore possible that the higher star formation rates in these objects could simply be due to them having had less time to deplete their newly-acquired gas reservoirs.

We alleviate this potential problem by requiring the dust lane control and sample galaxies to span the same range in starburst age, $100 < t_2 < 400$~Myr. As Fig.~\ref{fig:SFtimescales}(a) shows, this preferentially picks out the AGN population (as defined in the BPT plot), and should be representative of recent merging events in both the dust lane and overall early-type populations. We note that the stellar mass distributions of the two subsamples are very similar.

\begin{figure*}
\centering
\subfigure[star formation rate]{\includegraphics[height=0.3\textwidth,angle=0]{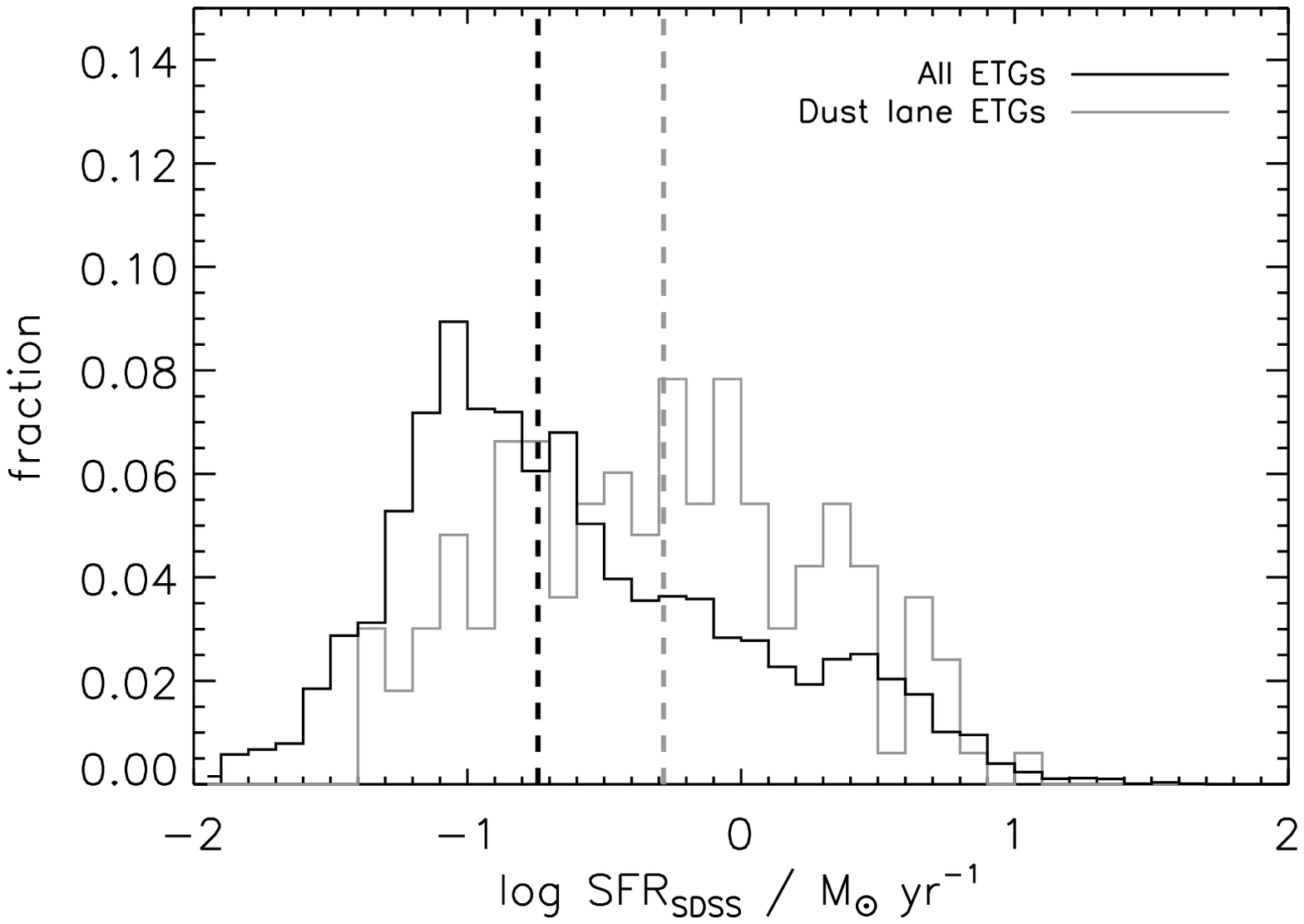}}
\subfigure[$\dot{m}_{\rm BH}$]{\includegraphics[height=0.3\textwidth,angle=0]{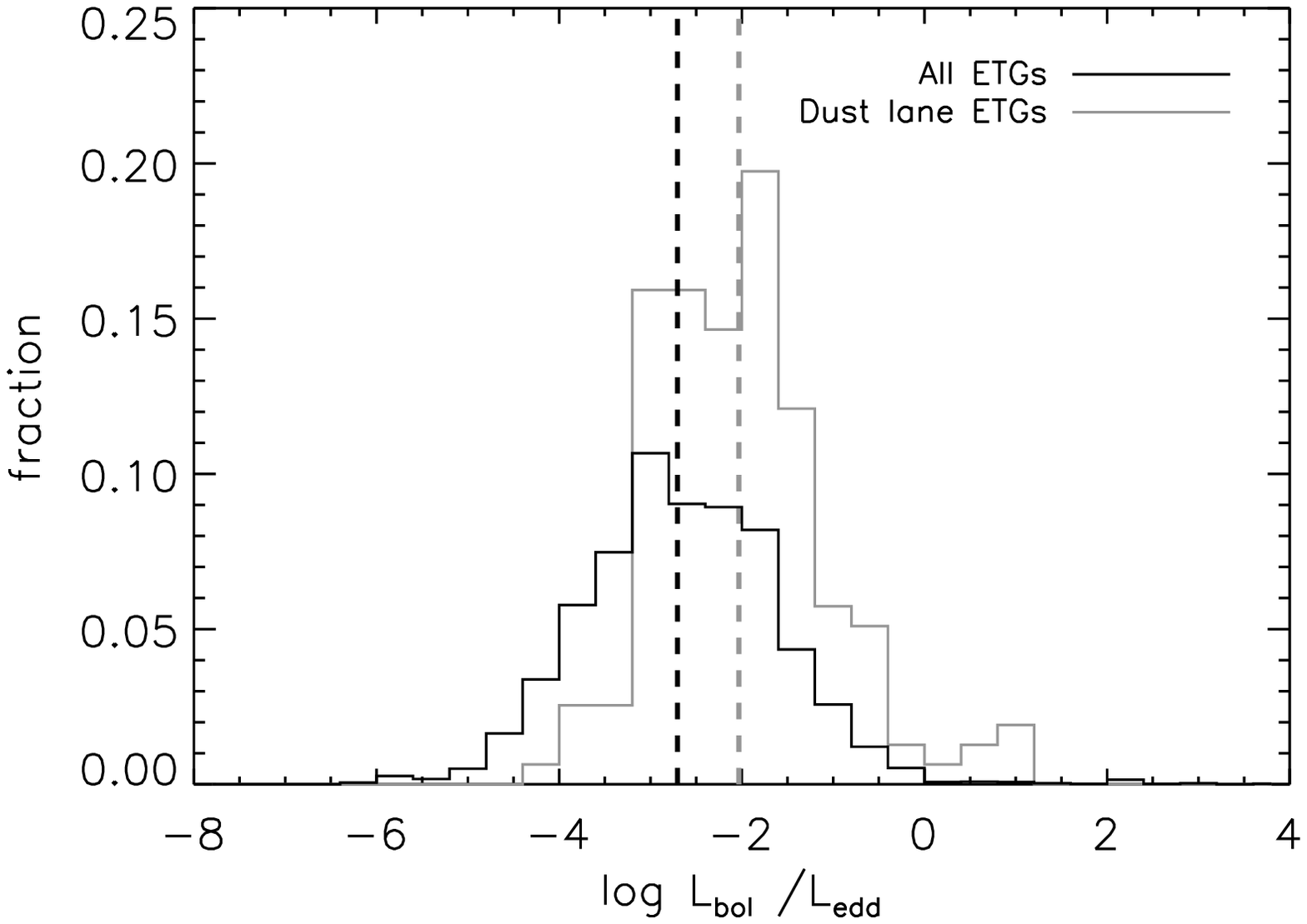}}
\caption{Left: distribution of star formation rates for dust lane early-type galaxies and a matched control sample of all early types. Both samples have $100 < t_2 < 400$~Myr. Right: distribution of $\dot{m}_{\rm BH}$ in Eddington units. Median values are again shown by dashed lines. Dust lane early-type galaxies have significantly higher star formation and black hole accretion rates than the control sample, consistent with these objects being associated with recent gas-rich mergers.}
\label{fig:gasAccretion}
\end{figure*}

Fig.~\ref{fig:gasAccretion}(a) shows that, even after correcting for the difference in starburst age distributions, dust lane galaxies exhibit star formation rates that are higher by a factor of 3 than the control sample. If dust corrections are important, this number is a lower limit, and the actual star formation rates in dust lane early-type galaxies may be even higher.

Fig.~\ref{fig:gasAccretion}(b) shows the distribution of the black hole accretion rate, in Eddington units. Bolometric luminosity is derived using the [O\,{\sc iii}] line luminosity, and the usual correction factor of 3500 \citep{hec04} is applied. The black hole mass is estimated via its velocity dispersion \citep{tre02}. Dust lane early types show accretion rates that are again a factor of three higher than the control sample, consistent with the discrepancy in the star formation rates.

A final confirmation of this picture comes from emission-line AGN fractions in the two samples. Almost all (84 per cent) dust lane early types are classified as either Seyfert, LINER or transition objects, compared to only 24 per cent for the control sample. More than 60 per cent of the objects in the control sample have emission lines that are too weak to place them on the BPT diagnostic, suggesting that these objects undergo very low-level (if any) star formation and AGN activity, despite having relatively young starbursts. This is in stark contrast to dust lane hosts with the same starburst ages, and confirms that the dust features are simply associated with galaxies that have acquired large amounts of gas, most likely in a recent merger.

\section{Conclusions}
\label{sec:conclusions}

We consider the origin and evolution of dust lane early-type galaxies. In a companion paper, we showed that these objects are preferentially located in poor environments, are morphologically disturbed, and have higher star formation rates than the mean for early-type galaxies.

In this work we addressed the issue of the origin of these objects. By employing optical (SDSS) and radio (FIRST and NVSS) surveys we examined AGN properties of dust lane early-type galaxies and a relevant control sample of early-type. Our findings are as follows.

\begin{enumerate}
\item[(1)] Dust lane early types are much more likely to host emission-line AGN than the control sample of early-type galaxies.
\item[(2)] Radio-loud AGN in dust lane early types do not show an environmental dependence, with similar fractions detected in the field and groups/clusters. By contrast, control sample early-type galaxies are much more likely to be radio loud if located in a cluster.
\item[(3)] Radio-loud AGN in dust lane early types are much more likely to also be detected in emission lines than radio-loud AGN in the control sample.
\end{enumerate}

These results all strongly suggest that minor mergers act as a trigger for AGN activity in dust lane early-type galaxies.

\begin{enumerate}
\item[(4)] Dust lane early types have starburst ages that are younger than the mean for early-type galaxies.
\item[(5)] Dust lane early-type galaxies exhibit higher star formation and black hole accretion rates than the control sample of early types with the same starburst ages.
\end{enumerate}

These results indicate that the minor mergers are both gas rich and recent. The mergers act as a trigger for both the starburst and AGN activity. The dust lane phase is associated with starburst galaxies in which the AGN has already been triggered, but has not yet shut off star formation completely.

\section*{Acknowledgements}

SSS thanks the Australian Research Council and New College, Oxford for research fellowships. YST is grateful to the French Ministry of Foreign and European Affairs for an Egide-Eiffel scholarship, and Ecole Polytechnique. SK acknowledges a Research Fellowship from the Royal Commission for the Exhibition of 1851, an Imperial College Junior Research Fellowship and a Senior Research Fellowship from Worcester College, Oxford.

We are grateful to Martin Bureau, Christophe Pichon and Adrianne Slyz for illuminating discussions, the referee for useful comments, and Sally Hales for assistance with FIRST and NVSS data.

This work would not have been possible without the contributions of citizen scientists as part of the Galaxy Zoo 2 project. 

Funding for the SDSS and SDSS-II has been provided by the Alfred P. Sloan Foundation, the Participating Institutions, the National Science Foundation, the US Department of Energy, the National Aeronautics and Space Administration, the Japanese Monbukagakusho, the Max Planck Society and the Higher Education Funding Council for England. The SDSS web site is http://www.sdss.org.

The SDSS is managed by the Astrophysical Research Consortium for the Participating Institutions. The Participating Institutions are the American Museum of Natural History, Astrophysical Institute Potsdam, University of Basel, University of Cambridge, Case Western Reserve University, University of Chicago, Drexel University, Fermilab, the Institute for Advanced Study, the Japan Participation Group, Johns Hopkins University, the Joint Institute for Nuclear Astrophysics, the Kavli Institute for Particle Astrophysics and Cosmology, the Korean Scientist Group, the Chinese Academy of Sciences (LAMOST), Los Alamos National Laboratory, the Max-Planck-Institute for Astronomy (MPIA), the Max-Planck-Institute for Astrophysics (MPA), New Mexico State University, Ohio State University, University of Pittsburgh, University of Portsmouth, Princeton University, the United States Naval Observatory and the University of Washington.

The NVSS and FIRST surveys were carried out using the National Radio Astronomy Observatory VLA. The National Radio Astronomy Observatory is a facility of the National Science Foundation operated under cooperative agreement by Associated Universities, Inc.

\end{document}